\documentclass[a4paper,11pt]{article}

\usepackage{jheppub} 
\usepackage[normalem]{ulem}  

\newcommand{\Ga}{{\Gamma}}
\newcommand{\De}{{\Delta}}

\newcommand{\al}{{\alpha}}
\newcommand{\bt}{{\beta}}
\newcommand{\ga}{{\gamma}}
\newcommand{\de}{{\delta}}
\newcommand{\ep}{{\epsilon}}

\newcommand{\te}{{\theta}}
\newcommand{\ka}{{\kappa}}

\newcommand{\bsb}{{\boldsymbol{b}}}
\newcommand{\bsk}{{\boldsymbol{k}}}

\newcommand{\bsx}{{\boldsymbol{x}}}
\newcommand{\bsy}{{\boldsymbol{y}}}

\newcommand{\bsep}{{\boldsymbol{\epsilon}}}
\newcommand{\bspd}{{\boldsymbol{\partial}}}

\newcommand{\bsK}{{\boldsymbol{K}}}

\newcommand{\cA}{{\mathcal{A}}}
\newcommand{\cB}{{\mathcal{B}}}
\newcommand{\cG}{{\mathcal{G}}}

\newcommand{\cI}{{\mathcal{I}}}
\newcommand{\cD}{{\mathcal{D}}}
\newcommand{\cK}{{\mathcal{K}}}

\newcommand{\fkD}{{\mathfrak{D}}}

\newcommand{\lan}{{\langle}}

\newcommand{\nn}{{\nonumber}}

\newcommand{\pd}{{\partial}}

\newcommand{\ran}{{\rangle}}

\newcommand{\wh}{\widehat}
\newcommand{\wt}{\widetilde}

\def\bbra{{\langle\kern-2.5pt\langle}}
\def\kket{{\rangle\kern-2.5pt\rangle}}

\newcommand{\cd}{\!\cdot\!}

\title{Momentum space approach to crossing symmetric CFT correlators}

\author[a]{Hiroshi Isono,}
\author[b]{Toshifumi Noumi}
\author[c]{and Gary Shiu}

\affiliation[a]{Department of Physics, Faculty of Science, Chulalongkorn University, Bangkok 10330, Thailand}
\affiliation[b]{Department of Physics, Kobe University, Kobe 657-8501, Japan}
\affiliation[c]{Department of Physics, University of Wisconsin-Madison, Madison, WI 53706, USA}

\emailAdd{hiroshi.isono81@gmail.com}
\emailAdd{tnoumi@phys.sci.kobe-u.ac.jp}
\emailAdd{shiu@physics.wisc.edu}

\preprint{KOBE-COSMO-18-06, MAD-TH-18-04}

\abstract{
We construct a crossing symmetric basis for conformal four-point functions in momentum space by requiring consistent factorization.
Just as scattering amplitudes factorize when the intermediate particle is on-shell, non-analytic parts of conformal correlators enjoy a similar factorization in momentum space.
Based on this property, Polyakov, in his pioneering 1974 work, introduced a basis for conformal correlators which manifestly satisfies the crossing symmetry. He then initiated the bootstrap program by requiring its consistency with the operator product expansion. This approach is complementary to the ordinary bootstrap program, which is based on the conformal block and requires the crossing symmetry as a consistency condition of the theory.
Even though Polyakov's original bootstrap approach has been revisited recently, the crossing symmetric basis has not been constructed explicitly in momentum space. In this paper we complete the construction of the crossing symmetric basis for scalar four-point functions with an intermediate operator with a general spin, by using new analytic expressions for three-point functions involving one tensor. Our new basis manifests the analytic properties of conformal correlators. Also the connected and disconnected correlators are manifestly separated, so that it will be useful for the study of large $N$ CFTs in particular.
}

\begin{document} 
\setcounter{tocdepth}{2}
\maketitle
\flushbottom

\section{Introduction}
\setcounter{equation}{0}

Conformal field theory (CFT) is a universal testbed for physicists and mathematicians. Starting from the scale invariance of high energy scatterings, it has experienced a vast range of applications, including critical phenomena and string theory. In particular, the outstanding development in $d=2$ CFT has brought about intensive interplay between physics and mathematics in connection with the representation theory, integrable systems and hypergeometric functions.
The AdS/CFT correspondence~\cite{Maldacena:1997re,Gubser:1998bc,Witten:1998qj} has established CFT as a probe of quantum gravity and triggered recent research activities in CFT on higher dimensional spacetimes (see, e.g.,~\cite{Qualls:2015qjb,Rychkov:2016iqz,Simmons-Duffin:2016gjk,Penedones:2016voo} for review articles and references therein). Moreover, its application reaches even cosmology, where the conformal group as the de Sitter isometry is crucial to analyze inflationary correlators~\cite{Maldacena:2002vr,Antoniadis:2011ib,Maldacena:2011nz,Creminelli:2012ed,Schalm:2012pi,Mata:2012bx,McFadden:2013ria,Ghosh:2014kba,Bzowski:2012ih,Kundu:2014gxa,Arkani-Hamed:2015bza,Kundu:2015xta,Shukla:2016bnu,Isono:2016yyj}. By further developing our CFT techniques, we would like to enlarge the scope of our theoretical toolkit.

\medskip
The basic ingredients of CFT are three-point functions. Once the operator spectrum is specified, four-point and higher order correlators can in principle be calculated from three-point functions by using the operator product expansion (OPE). For example, four-point functions of primary scalars with an identical conformal mass dimension $\Delta$ are given by
\begin{align}
\lan O_1(\bsx_1)O_2(\bsx_2)O_3(\bsx_3)O_4(\bsx_4) \ran =
\sum_{n}\widetilde{C}_{12n}\widetilde{C}_{34n}x_{12}^{-2\Delta}x_{34}^{-2\Delta}
G_{n}\left(\frac{x_{12}^2x_{34}^2}{x_{13}^2x_{24}^2},\frac{x_{14}^2x_{23}^2}{x_{13}^2x_{24}^2}\right)
\,,
\label{4pt-confblock}
\end{align}
where the summation is over all the primary operators and the numerical constants $\widetilde{C}_{ijk}$ are the OPE coefficients. The conformal block $G_{n}$ can be determined uniquely by conformal symmetry. In this expansion, the crossing symmetry is not manifest, hence it gives a nontrivial constraint on the theory together with unitarity or reflection positivity (if any)~\cite{Ferrara:1973yt}.
This conformal bootstrap approach has been very successful in $d=2$~\cite{Belavin:1984vu}. Furthermore, recent years have seen a remarkable progress in its application to CFTs in higher dimensions. Following the seminal works by Dolan and Osborn~\cite{Dolan:2000ut,Dolan:2003hv,Dolan:2011dv}, various properties of conformal blocks have been clarified so far and the bootstrap program has been promoted both numerically and analytically~\cite{Penedones:2010ue,Fitzpatrick:2011ia,Costa:2011dw,ElShowk:2012ht,Hogervorst:2013sma,Hogervorst:2013kva,Kos:2013tga,Gliozzi:2013ysa,Gliozzi:2014jsa,Hijano:2015zsa,Isachenkov:2016gim,Alday:2016njk,Caron-Huot:2017vep,Alday:2017zzv}.

\medskip
While the recent development in the bootstrap program is mostly based on the ordinary conformal block, Polyakov in his pioneering 1974 work~\cite{Polyakov:1974gs} employed another basis for conformal correlators as a framework for the bootstrap program. Analogously to the K\"all\'en-Lehmann spectral representation of scattering amplitudes, he introduced a crossing symmetric basis for momentum space correlators as
\begin{align}
\langle O_1(\bsk_1)O_2(\bsk_2)O_3(\bsk_3)O_4(\bsk_4)\rangle'
=\sum_n\left(W_n^{(\mathbf{s})}+W_n^{(\mathbf{t})}+W_n^{(\mathbf{u})}\right)\,,
\end{align}
where we introduced $\langle\,\ldots\,\rangle=(2\pi)^d\delta^d(\sum \bsk_i)\langle\,\ldots\,\rangle'$ and
 $W_n^{(\mathbf{s})}$ is what we call the $\mathbf{s}$-channel Polyakov block enjoying a consistent $\mathbf{s}$-channel factorization as we elaborate in Sec.~\ref{subsec:Polyakov}. In this basis, the crossing symmetry and thus (non-)analyticity in each channel are manifest by construction. On the other hand, consistency with the OPE is obscured, hence it gives a nontrivial constraint on the theory.
Polyakov demonstrated that the consistency with the OPE determines anomalous dimensions and OPE coefficients of $O(N)$ models in the $d=4-\epsilon$ dimension (see~\cite{Sen:2015doa} for a recent generalization). This bootstrap approach therefore plays a complementary role to the one based on the ordinary conformal block.

\medskip
Even though Polyakov's original bootstrap approach is very elegant, this direction has not been explored very much until recently, essentially because of technical complications due to conformal symmetry in momentum space. Indeed, Polyakov did not derive an explicit form of the block $W_n^{(\mathbf{s})}$ in momentum space, but rather he proposed a crossing symmetric basis in position space. Recently in~\cite{Gopakumar:2016wkt,Gopakumar:2016cpb}, Gopakumar et al. revisited Polyakov's bootstrap approach in Mellin space. In particular, they showed that the crossing symmetric basis is nothing but the Witten exchange diagram by explicit computations in Mellin space. To further promote the recent revival of Polyakov's original bootstrap idea, we would like to complete the construction of the crossing symmetric basis in momentum space in light of recent developments on conformal correlators in momentum space.

\begin{figure}
\begin{center}
\includegraphics[width=115mm, bb=0 0 387 157]{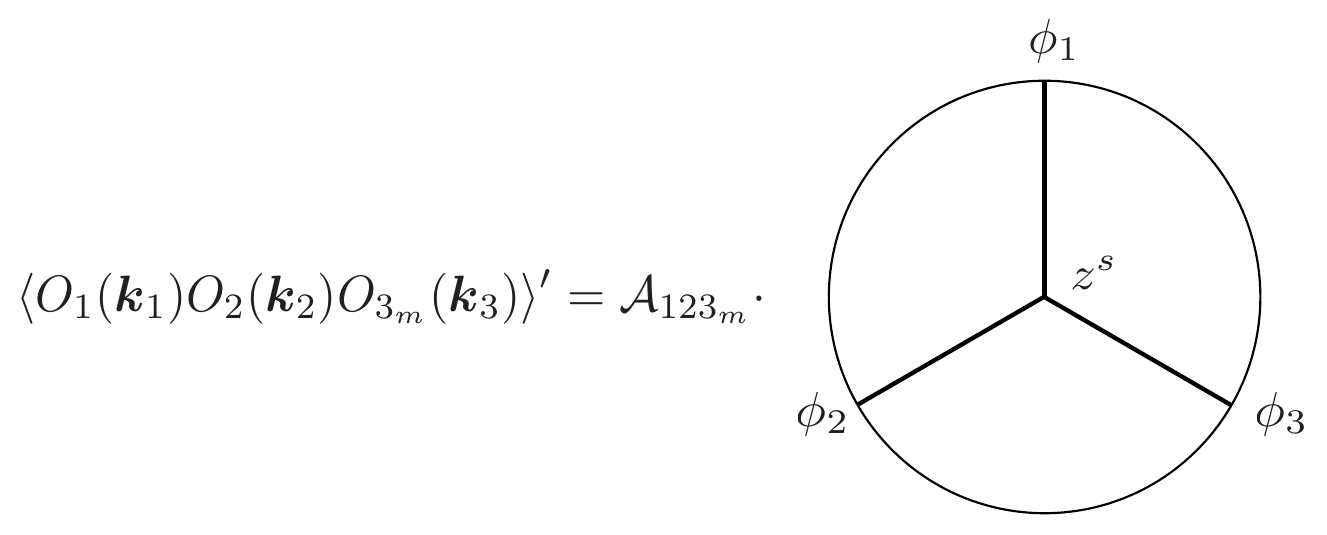}
\end{center}
\vspace{-5mm}
\caption{Three-point functions of two scalars and one tensor can be obtained by acting a differential operator $\cA_{123_m}$ on the cubic Witten diagram of three scalars, with an additional factor $z^s$ multiplied to the vertex in the integrand.}
\label{fig:3pt_Witten_helicity}
\end{figure}

\medskip
The purpose of this paper is to explicitly construct the crossing symmetric basis in momentum space, following Polyakov's original argument based on analyticity and factorization. As a first step, in the present paper, we focus on scalar four-point functions in $d=3$ Euclidean space.\footnote{The results in general dimensions will be presented in~\cite{crossing2}.} Since our work is rather technical, we would like to outline the punchline of our results in the rest of this introduction.
For our construction, it is important to first clarify the analytic properties of three-point functions.
Conformal three-point functions in momentum space were intensively studied recently in~\cite{Ferrara:1974nf,Bzowski:2012ih,Bzowski:2013sza,Bzowski:2015pba,Bzowski:2015yxv,Coriano:2012hd,Coriano:2013jba,Bzowski:2017poo,Kundu:2014gxa,Arkani-Hamed:2015bza,Kundu:2015xta,Shukla:2016bnu,Coriano:2018Feb}.
For example, three-point functions of primary scalars are given by~\cite{Ferrara:1974nf,Bzowski:2012ih,Bzowski:2013sza,Bzowski:2015pba,Bzowski:2015yxv}
\begin{align}
\label{scalar_3pt_intro}
\langle O_1(\bsk_1)O_2(\bsk_2)O_3(\bsk_3)\rangle'
=C_{123}
\int_0^\infty \frac{dz}{z^{4}}\mathcal{B}_{\nu_1}(k_1;z)\mathcal{B}_{\nu_2}(k_2;z)\mathcal{B}_{\nu_3}(k_3;z)\,,
\end{align}
where $C_{123}$ is the OPE coefficient and $\mathcal{B}_{\nu}$ is the bulk-to-boundary propagator of the (would-be) dual bulk scalar. Interestingly, the radial coordinate $z$ of the AdS Poincar\'e patch naturally appears when one solves the conformal Ward-Takahashi identity in momentum space. We emphasize that our argument does not rely on the AdS/CFT correspondence, even though we use its terminology to make the notation intuitive. As a consequence, analytic properties of three-point functions are captured by those of the bulk-to-boundary propagators. For construction of the Polyakov block for an intermediate spinning operator, we need three-point functions of two scalars and one tensor as well. As depicted in Fig.~\ref{fig:3pt_Witten_helicity}, we find in Appendix~\ref{App-sec:3pt} that they are obtained by simply acting a differential operator $\cA_{123_m}$ (its definition is given in Sec.~\ref{subsec:spin_3pt}) on an integral similar to Eq.~\eqref{scalar_3pt_intro} as\footnote{
The integral~\eqref{scalar-scalar-tensor_intro} is convergent only when $|{\rm Re}\,\nu_1|+|{\rm Re}\,\nu_2|+|{\rm Re}\,\nu_3|<s+\frac{3}{2}$. Otherwise, there appears a singularity near $z=0$ and we need to perform analytic continuation~\cite{Bzowski:2015pba,Bzowski:2015yxv}, which may be carried out,
e.g., by introducing the Pochhammer contour. A similar remark is applicable to
 our integral representation for the Polyakov block.}
\begin{align}
\label{scalar-scalar-tensor_intro}
\lan O_1(\bsk_1)O_2(\bsk_2)O_{3_m}(\bsk_3) \ran'
&= \cA_{123_m}\int_0^\infty \frac{dz}{z^{4}}z^s\mathcal{B}_{\nu_1}(k_1;z)\mathcal{B}_{\nu_2}(k_2;z)\mathcal{B}_{\nu_3}(k_3;z)
\,,
\end{align}
where we introduced what we call the helicity basis by analogy with helicities of massless particles, and $O_{n_m}$ has a spin $s$ and a helicity $m$~\cite{Arkani-Hamed:2015bza}. As we discuss in Sec.~\ref{subsec:spin_analytic}, the differential operator $\cA_{123_m}$ is analytic in momenta, so that all the non-analyticities of three-point functions are captured by the bulk-to-boundary propagator $\mathcal{B}_{\nu}(k;z)$ in the integrand. This is a technically important observation for our construction.

\begin{figure}
\begin{center}
\includegraphics[width=90mm, bb=0 0 579 267]{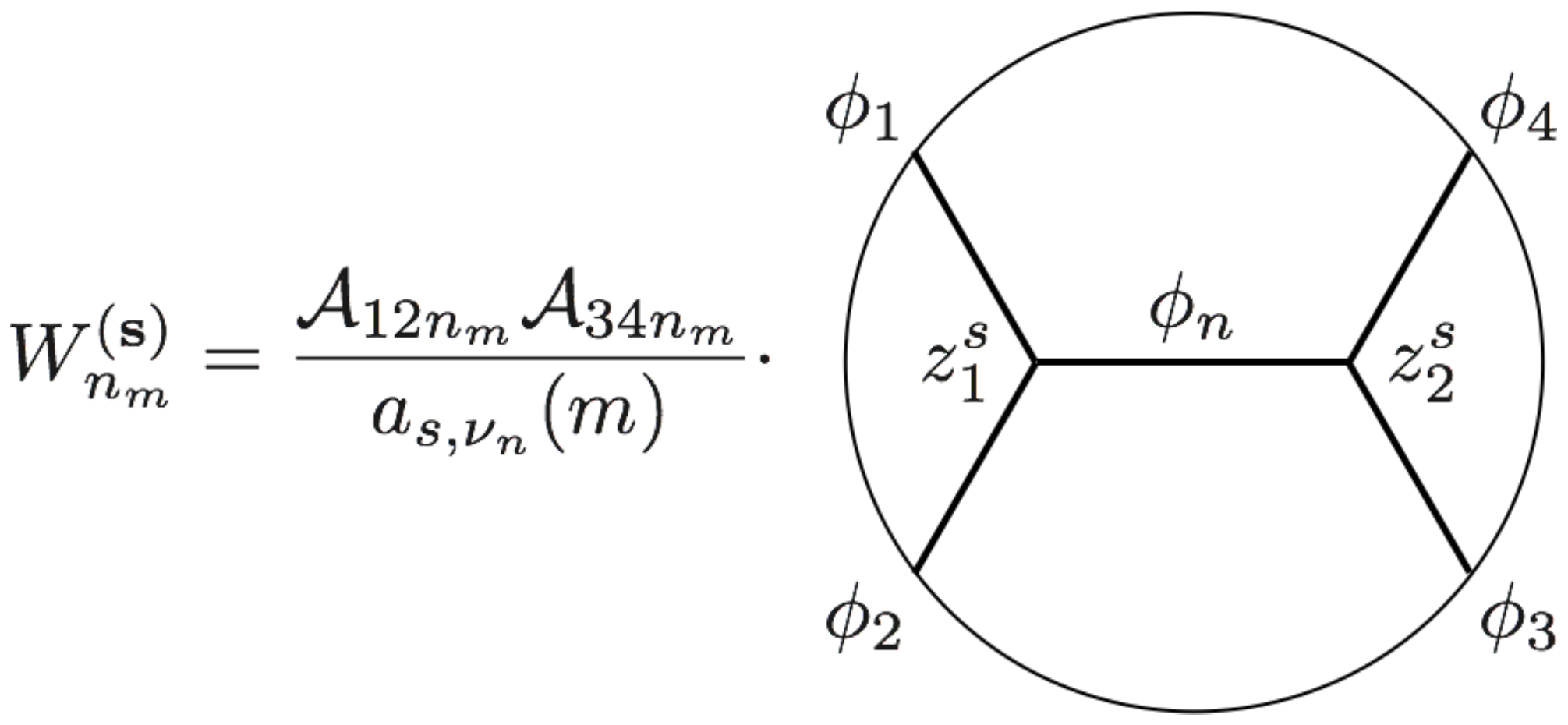}
\end{center}
\vspace{-5mm}
\caption{The Polyakov block for an intermediate spinning operator can be obtained by acting two differential operators $\cA_{12n_m}$ and $\cA_{34n_m}$ on the scalar Witten exchange diagram with an additional factor $z^s$ multiplied to the cubic vertices in the integrand.}
\label{fig:Polyakov_helicity}
\end{figure}

\medskip
The above analytic properties of three-point functions now enable us to construct the crossing symmetric basis. As we explain in Sec.~\ref{subsec:Polyakov}, Polyakov's idea is to require that non-analytic parts of the Polyakov block $W_n^{(\mathbf{s})}$ enjoy a consistent factorization, just as the on-shell factorization of scattering amplitudes. In Sec.~\ref{subsec:Witten}, we show that the Polyakov block for an intermediate scalar is nothing but the Witten exchange diagram:
\begin{align}
W_n^{(\mathbf{s})}&=C_{12n}C_{34n}\int_0^\infty \frac{dz_1}{z_1^{4}}\int_0^\infty \frac{dz_2}{z_2^{4}}
\nn\\*
\label{scalar-exchange_intro}
&\quad
\times
\mathcal{B}_{\nu_1}(k_1;z_1)\mathcal{B}_{\nu_2}(k_2;z_1)
\cG_{\nu_n}(k_{12};z_1,z_2)
\mathcal{B}_{\nu_3}(k_3;z_2)\mathcal{B}_{\nu_4}(k_4;z_2)\,,
\end{align}
where $\cG_{\nu_n}(k_{12};z_1,z_2)$ is the bulk-to-bulk propagator. The essential point here is that the $z$-ordered part of the bulk-to-bulk propagator is analytic in $k_{12}$. As a result, the non-analytic parts associated to the intermediate on-shell state manifestly reproduce the factorization required by the Polyakov ansatz. This shows that momentum space is a natural language for this program. It is then not difficult to generalize the construction to intermediate spinning operators. The Polyakov block $W_{n_m}^{(\mathbf{s})}$ for an intermediate operator $O_{n_m}$ with a spin $s$ and a helicity $m$ can be obtained again by acting the differential operators $\cA_{12n_m}$ and $\cA_{34n_m}$ on an integral similar to Eq.~\eqref{scalar-exchange_intro} as (see also Fig.~\ref{fig:Polyakov_helicity})
\begin{align}
W_{n_m}^{(\mathbf{s})}&=\frac{\cA_{12n_m}\cA_{34n_m}}{a_{s,\nu_n}(m)}
\int_0^\infty
\frac{dz_1}{z_1^{4}}
\int_0^\infty
\frac{dz_2}{z_2^{4}}z_1^sz_2^s
\nn\\*
&\quad
\times
\mathcal{B}_{\nu_1}(k_1;z_1)\mathcal{B}_{\nu_2}(k_2;z_1)
\cG_{\nu_n}(k_{12};z_1,z_2)
\mathcal{B}_{\nu_3}(k_3;z_2)\mathcal{B}_{\nu_4}(k_4;z_2)\,,
\end{align}
where $a_{s,\nu_n}(m)$ is a numerical factor associated to the normalization of $O_{n_m}$. This is our main result.
As a side remark, we would like to point out that the connected and disconnected correlators are manifestly separated in this basis (see Sec.~\ref{subsec:identity}). We therefore believe that our basis is useful not only for the Polyakov type bootstrap, but also for the study of large $N$ CFTs, which are holographically dual to weakly coupled bulk theories.

\bigskip
The rest of this paper is organized as follows. In Sec.~\ref{sec:scalar23} we discuss analytic properties of scalar two- and three-point functions in momentum space. In particular we demonstrate that non-analytic parts of three-point functions factorize into analytic cubic vertices and two-point functions. In Sec.~\ref{Sec:intermediate_scalar} we introduce Polyakov's analyticity based argument and illustrate that momentum space provides a natural language for this program. By requiring consistent factorization in each channel, we show that the Polyakov blocks for intermediate scalars are nothing but the Witten exchange diagrams. In Sec.~\ref{sec:spin} we then generalize the argument to the intermediate spinning operators and complete the construction of the crossing symmetric basis. Some technical details are collected in Appendices.

\section{Analytic properties of scalar two- and three-point functions}
\setcounter{equation}{0}
\label{sec:scalar23}

In this section we discuss analytic properties of two- and three-point functions in momentum space. To avoid technical complication associated with spins, we first focus on scalar correlators in this section. As we will see in Sec.~\ref{sec:spin}, however, most of analytic properties discussed in this section are carried over to correlators involving operators with spins. Throughout the paper, we work on the $d=3$ Euclidean space unless otherwise stated.

\medskip
In the first two subsections we review basics of two- and three-point functions in momentum space. In particular we introduce a building block of three-point functions called the triple-$K$ integral~\cite{Bzowski:2013sza,Bzowski:2015pba,Bzowski:2015yxv}, which is essentially the cubic Witten diagram. We then elaborate on its analytic properties in the last subsection. The cubic vertices introduced there are relevant when constructing a crossing symmetric basis for conformal correlators.

\subsection{Two-point functions}

In momentum space two-point functions of primary scalars are given by
\begin{align}
\label{scalar_2pt_momentum}
\langle O(\bsk)O(-\bsk)\rangle'=C_{OO}\left(k^2\right)^{\Delta-3/2}
=C_{OO}\left(k^2\right)^{\nu}\,,
\end{align}
where $k=|\bsk|$, $\Delta$ is the scaling dimension of $O$, and $\nu=\Delta-3/2$. In this paper, for notational simplicity, we use $\langle\,\ldots\,\rangle'$ for correlation functions with the momentum conservation factor dropped. More explicitly, $\langle\,\ldots\,\rangle=(2\pi)^3\delta^3(\sum \bsk_i)\langle\,\ldots\,\rangle'$. The corresponding two-point functions in position space are
\begin{align}
\langle O(\bsx_1)O(\bsx_2)\rangle=\widetilde{C}_{OO}\left(x_{12}^2\right)^{-\Delta}
\quad
{\rm with}
\quad
\widetilde{C}_{OO}=2^{2\nu} \pi^{-3/2}\frac{\Ga(\nu+3/2)}{\Ga(-\nu)}C_{OO}
\,,
\end{align}
where notice that $\widetilde{C}_{OO}$ is positive in unitary CFTs, but $C_{OO}$ is not necessarily positive.

\medskip
As is clear from Eq.~\eqref{scalar_2pt_momentum}, two-point functions are not analytic in $\bsk$, or equivalently in $k^2$, unless $\nu$ is an integer.\footnote{For simplicity, we do not consider operators with an integer $\nu$ in this paper.} When we analytically continue $k^2$, we introduce a  branch cut along $k^2<0$. The discontinuity on the branch cut is
\begin{align}
\text{Disc}_{k^2} \,\langle O(\bsk)O(-\bsk)\rangle'
&=C_{OO}\left|k^2\right|^{\nu}\left(e^{i\pi\nu}-e^{-i\pi\nu}\right)=2i\,\text{Im}\,\langle O(\bsk)O(-\bsk)\rangle'\,,
\end{align}
where the imaginary part of the two-point function on the r.h.s. is evaluated at $\text{Im}\,k^2=\epsilon$ with a positive infinitesimal number $\epsilon$ as usual.

\subsection{Three-point functions}

In momentum space three-point functions of primary scalars are given by~\cite{Ferrara:1974nf,Bzowski:2013sza,Bzowski:2015pba,Bzowski:2015yxv}
\begin{align}
\label{triple-K}
\langle O_1(\bsk_1)O_2(\bsk_2)O_3(\bsk_3)\rangle'
=C_{123}
\int_0^\infty \frac{dz}{z^{4}}\mathcal{B}_{\nu_1}(k_1;z)\mathcal{B}_{\nu_2}(k_2;z)\mathcal{B}_{\nu_3}(k_3;z)\,,
\end{align}
where $\mathcal{B}_{\nu}$ is the bulk-to-boundary propagator of the (would-be) dual bulk scalar:\footnote{
We emphasize that our argument does not rely on the AdS/CFT correspondence. We, however, use the term ``bulk-to-boundary propagator" of the {\it would-be} dual bulk field to make the notation more intuitive. Also, for notational simplicity, we call the {\it would-be} dual field simply the dual field in the rest of this paper.}
\begin{align}
\label{bulk_boundary_K}
\mathcal{B}_\nu(k;z)&=\frac{1}{2^{\nu-1}\Gamma(\nu)}k^\nu z^{3/2}K_\nu(kz)
\\
\label{bulk_boundary_I}
&=\frac{\Gamma(1-\nu)}{2^\nu}z^{3/2-\nu}\Big[(kz)^{\nu}I_{-\nu}(kz)-(kz)^{\nu}I_{\nu}(kz)\Big]\,.
\end{align}
Also $I_\nu(z)$ and $K_\nu(z)$ are the modified Bessel functions of the first and the second kinds, respectively. We call them the Bessel $I$ and $K$ functions in short. The bulk-to-boundary propagator $\mathcal{B}_\nu(k;z)$ is a solution for the scalar equation of motion on $AdS_4$,
\begin{align}
\label{b-to-bd_def}
\Big[z^2\partial_z^2-2z\partial_z-z^2k^2-m^2\Big]\mathcal{B}_\nu(k;z)=0\,,
\end{align}
where $m^2=\nu^2-9/4$ is the dual scalar mass. Our normalization is chosen such that
\begin{align}
\label{expansion_B}
\mathcal{B}_\nu(k;z)\to z^{3/2-\nu}\Big(1+\mathcal{O}\big((kz)^2\big)\Big)+\frac{\Gamma(-\nu)}{2^{2\nu}\Gamma(\nu)}k^{2\nu}z^{3/2+\nu}\Big(1+\mathcal{O}\big((kz)^2\big)\Big)\,.
\end{align}
As is clear from the large $z$ behavior of modified Bessel functions,
\begin{align}
\label{KI_large_z}
K_\nu(z)\to\sqrt{\frac{\pi}{2}}z^{-1/2}e^{-z}\,,
\quad
I_\nu(z)\to \frac{1}{\sqrt{2\pi}}z^{-1/2}e^z
\quad
{\rm for}
\quad
z\gg1\,,
\end{align}
the bulk-to-boundary propagator $\mathcal{B}_\nu(k;z)$ decays in the limit $kz\to\infty$. Since the integral~\eqref{triple-K} contains three $K_\nu(z)$'s, it is called the triple-$K$ integral. We refer the reader to the references~\cite{Bzowski:2015pba,Bzowski:2015yxv} for its detailed properties.

\medskip
The origin of the triple-$K$ integral may be understood in the following three ways:
\begin{enumerate}
\item First, the integral~\eqref{triple-K} is nothing but the Witten diagram for the cubic coupling $\phi_1\phi_2\phi_3$ ($\phi_i$ is the dual of $O_i$) if we identify $z$ with the radial coordinate of the Poincar\'e patch. It respects the AdS isometry by construction, hence it is conformally covariant.

\item Indeed, we may explicitly check that the triple-$K$ integral satisfies the conformal Ward-Takahashi identity in momentum space. By requiring an appropriate analytic behavior, scalar three-point functions may be determined uniquely up to an overall coefficient from conformal symmetry~\cite{Bzowski:2015pba}.

\item It is also possible to obtain the triple-$K$ integral by Fourier transforming the position space three-point function,
\begin{align}
\lan O_1(\bsx_1)O_2(\bsx_2)O_3(\bsx_3) \ran 
= \frac{\widetilde{C}_{123}}{(x_{12}^2)^{\frac{1}{2}\De_{123}}
(x_{13}^2)^{\frac{1}{2}\De_{132}}(x_{23}^2)^{\frac{1}{2}\De_{231}}}
\,,
\end{align}
where $\De_{ijk}=\De_i+\De_j-\De_k$. As we compute in Appendix~\ref{App-subsec:3pt-FT}, the OPE coefficient $\widetilde{C}_{123}$ in position space is related to the one $C_{123}$ in momentum space as
\begin{align}
\widetilde{C}_{123}=\frac
{\Ga(\frac{\De_{231}}{2}) \Ga(\frac{\De_{132}}{2}) \Ga(\frac{\De_{123}}{2}) 
\Ga(\frac{\De_1+\De_2+\De_3-3}{2})}
{2 \pi^{3} \Gamma(\nu_1)\Gamma(\nu_2)\Gamma(\nu_3)}
C_{123}\,.
\end{align}
\end{enumerate}

\subsection{Factorization of three-point functions}
\label{subsec:factorization3pt}

\begin{figure}[t]
\begin{center}
\includegraphics[width=120mm, bb=0 0 341 106]{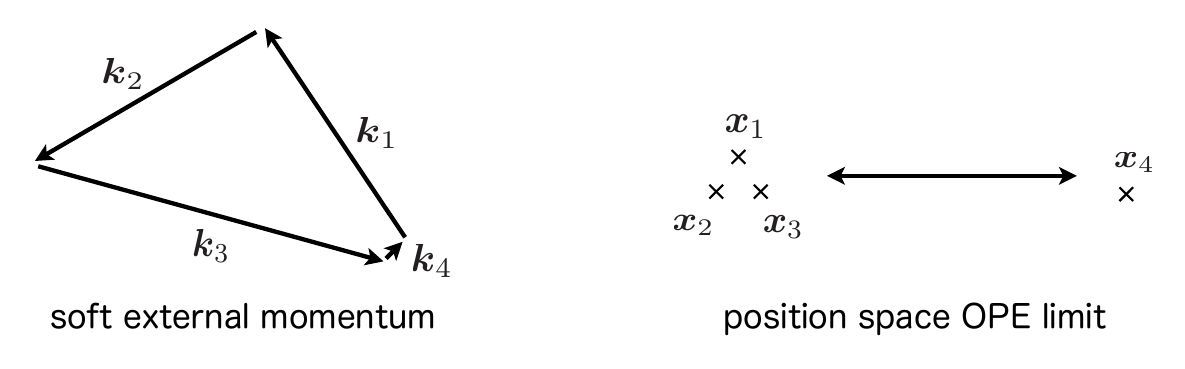}
\end{center}
\vspace{-5mm}
\caption{Due to momentum conservation, external momenta of correlation functions form a polygon. The soft limit of an external momentum, say $k_4\to0$ (the left figure), corresponds to the OPE limit in position space, where one operator is far separated from the others (the right figure).}
\label{fig:squeezed_OPE}
\end{figure}

We then discuss analyticity of three-point functions. First, the triple-$K$ integral~\eqref{triple-K} is defined in such a way that it is regular after an appropriate analytic continuation in $\nu_i$. It is therefore analytic as long as the integrand is analytic. Non-analyticity then shows up only in the squeezed limit, where one of the three momenta is much smaller than the other two. As the soft limit expansion~\eqref{expansion_B} of the bulk-to-boundary propagator shows, the only source of non-analyticity is the second term in the square bracket of Eq.~\eqref{bulk_boundary_I}. Note that the soft limit of external momenta is nothing but the momentum space counterpart of the OPE limit in position space (see Fig.~\ref{fig:squeezed_OPE}). The non-analytic component is then responsible for the intermediate on-shell state and thus the singular behavior in the OPE limit.

\medskip
Let us take a closer look at analyticity of three-point functions $\langle O_1(\bsk_1)O_2(\bsk_2)O_3(\bsk_3)\rangle'$ with respect to $\bsk_3$ (with $\bsk_1$ fixed). As we mentioned, the only source of non-analyticity is the bulk-to-boundary propagator for $O_3$.
The non-analytic part of three-point functions then factorizes into analytic cubic vertices and the discontinuity (imaginary part) of two-point functions as
\begin{align}
\nonumber
\text{Disc}_{k_3^2}\langle O_1(\bsk_1)O_2(\bsk_2)O_3(\bsk_3)\rangle'
&=T_{12;3}(\bsk_1,\bsk_2;\bsk_3)C_{O_3O_3}\text{Disc}_{k_3^2}\left(k_3^2\right)^{\nu_3}
\\*
&=T_{12;3}(\bsk_1,\bsk_2;\bsk_3)\,\text{Disc}_{k_3^2}\langle O_3(\bsk_3)O_3(-\bsk_3)\rangle'\,.
\end{align}
Here we defined the cubic vertex $T_{12;3}$ by
\begin{align}
&T_{12;3}(\bsk_1,\bsk_2;\bsk_3)
\nn\\*
\label{cubic_V}
&=-\frac{\Gamma(1-\nu_3)}{2^{\nu_3}}\frac{C_{123}}{C_{O_3O_3}}
\int_0^\infty \frac{dz}{z^{4}}\mathcal{B}_{\nu_1}(k_1;z)\mathcal{B}_{\nu_2}(k_2;z)z^{3/2}k_3^{-\nu_3}I_{\nu_3}(k_3z)\,,
\end{align}
which is analytic with respect to $\bsk_3$ around $\bsk_3=0$. We use this cubic vertex in the next section when constructing the crossing symmetric basis for conformal four-point functions.

\medskip
For later use, it is convenient to notice that the cubic vertex~\eqref{cubic_V} has a singularity in the collinear limit defined by $k_1+k_2=k_3$: As we mentioned, the Bessel $K$ and $I$ functions behaves in the large $z$ region as Eq.~\eqref{KI_large_z},
so that the bulk-to-boundary propagator has an exponential suppression $\sim e^{-kz}$. Since the integrand of cubic vertices contains the Bessel $I$ function, it has an exponential factor $\sim e^{-(k_1+k_2-k_3)z}$ in the large $z$ region. The integral then converges only when $k_1+k_2>k_3$. In particular, the exponential suppression disappears in the collinear limit $k_1+k_2=k_3$ and thus
becomes singular and non-analytic.

\section{Crossing symmetric basis: intermediate scalar}
\setcounter{equation}{0}
\label{Sec:intermediate_scalar}

In this section we construct a crossing symmetric basis for momentum space four-point functions,
following Polyakov's argument based on analyticity and factorization~\cite{Polyakov:1974gs}. For technical simplicity, we focus on the case with an intermediate scalar in this section. Extension to the case with an intermediate spinning operator will be given in the next section.

\medskip
After introducing Polyakov's argument in Sec.~\ref{subsec:Polyakov}, we show in Sec.~\ref{subsec:Witten} that the crossing symmetric basis is nothing but the Witten exchange diagram. Actually, this equivalence was already shown in Mellin space by explicit computations~\cite{Gopakumar:2016cpb}. However, we find that momentum space provides a natural language for this program.  In particular we elaborate on the emergence of the bulk-to-bulk propagator by studying analytic properties in momentum space. Sec.~\ref{subsec:identity} is devoted to the intermediate identity operator case, which is separately treated in our construction. We there emphasize that the connected and disconnected contributions to four-point functions are manifestly separated in this approach.

\subsection{Factorization ansatz \`a la Polyakov}
\label{subsec:Polyakov}

In ordinary Lorentzian field theories, scattering amplitudes are analytic off shell and their non-analytic parts, associated to on-shell particle creations, enjoy the factorization property. For example, tree-level four-point amplitudes in $\phi^3$ theory are schematically given~by
\begin{align}
\text{(4pt scattering)}=\text{(3pt vertex)} \times\frac{1}{m^2-\mathbf{s}-i\epsilon}\times\text{(3pt vertex)}+\text{($\mathbf{t},\mathbf{u}$-channels)}\,,
\end{align}
where $\mathbf{s}=-(k_1+k_2)^2$ is the Mandelstam variable. When the exchanged particle is on-shell, the propagator becomes non-analytic and the non-analytic part factorizes as
\begin{align}
{\rm Disc}_\mathbf{s}\,[\text{4pt scattering}]=\text{(3pt vertex)} \times{\rm Disc}_{\mathbf{s}}\left[\frac{1}{m^2-\mathbf{s}-i\epsilon}\right]\times\text{(3pt vertex)}\,.
\end{align}
Notice that cubic vertices and $\mathbf{t},\mathbf{u}$-channel diagrams are analytic in $\mathbf{s}$ since our interaction is local. Also, contact vertices, if any, are analytic as long as we consider local interactions.

\begin{figure}[t]
\begin{center}
\includegraphics[width=120mm, bb=0 0 353 82]{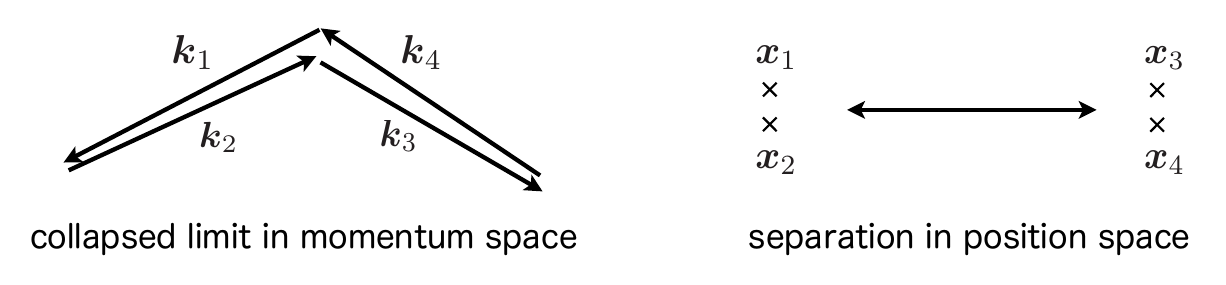}
\end{center}
\vspace{-5mm}
\caption{In the collapsed limit, an internal momentum, say $\bsk_{12}$, becomes soft compared to external momenta (the left figure). This limit corresponds to the OPE limit in position space, where operators are separated into two groups far away as depicted in the right figure.}
\label{fig:collapsed_OPE}
\end{figure}

\medskip
Let us now move on to our CFT argument.
In~\cite{Polyakov:1974gs} Polyakov introduced a CFT correlator analogue of such a factorization property.
As we demonstrated in Sec.~\ref{sec:scalar23}, non-analytic parts of three-point functions can be reformulated as
\begin{align}
{\rm Disc}_{k_3^2} \lan O_1(\bsk_1) O_2(\bsk_2) O_3(\bsk_3) \ran' 
= T_{12;3}(\bsk_1,\bsk_2;\bsk_3) \, {\rm Disc}_{k_3^2}\lan O_3(-\bsk_3)O_3(\bsk_3) \ran'\,.
\end{align}
Notice that factorization occurs even in three-point functions because we are considering correlators rather than scattering amplitudes. Note also that $T_{12;3}$ is analytic in $\bsk_3$ around $\bsk_3=\boldsymbol{0}$. In analogy with ordinary field theories, we call $T_{12;3}$ the cubic vertex.\footnote{To be precise, $T_{12;3}$ is the cubic vertex multiplied by two-point functions of $O_1$ and $O_2$. Indeed, it is non-analytic with respect to $\bsk_1$ and $\bsk_2$. We, however, call it the cubic vertex for simplicity.} Similarly, we require that non-analytic parts of four-point functions are factorized into the form,
\begin{align}
\nonumber
&{\rm Disc}_{\mathbf{s}}\langle O_1(\bsk_1)O_2(\bsk_2)O_3(\bsk_3)O_4(\bsk_4)\rangle'
\\
\label{s_factorization}
&=\sum_nT_{12;n}(\bsk_1,\bsk_2;-\bsk_{12})\,\text{Disc}_{\mathbf{s}}\langle O_n(\bsk_{12})O_n(-\bsk_{12})\rangle'T_{34;n}(\bsk_3,\bsk_4;-\bsk_{34})\,,
\end{align}
where the intermediate operator $O_n$ runs over all the primary operators. They are described in the basis with diagonal two-point functions in particular. We also introduced  $\bsk_{ij}=\bsk_i+\bsk_j$. Later, we use the Mandelstam type variables $\mathbf{s},\mathbf{t}$, and $\mathbf{u}$ defined by
\begin{align}
\mathbf{s}
=-(\bsk_1+\bsk_2)^2\,,
\quad
\mathbf{t}=-(\bsk_1+\bsk_3)^2\,,
\quad
\mathbf{u}=-(\bsk_1+\bsk_4)^2\,.
\end{align}
Since the on-shell conditions are not imposed on the external momenta, these variables are independent in contrast to the scattering amplitude case. Notice here that the collapsed limit $\bsk_{ij}\to\boldsymbol{0}$, where the non-analyticity shows up, is the momentum space analogue of the position space OPE limit (see Fig.~\ref{fig:collapsed_OPE}). In a similar fashion, we require factorization in the $\mathbf{t},\mathbf{u}$-channels as well.

\medskip
Based on the factorization property given above, Polyakov introduced a crossing symmetric basis for conformal four-point functions as\footnote{
Just as in the case of the recursion relation for scattering amplitudes~\cite{Elvang:2013cua}, there may exist analytic terms which cannot be determined only from the analyticity argument, even though earlier works~\cite{Polyakov:1974gs,Sen:2015doa,Gopakumar:2016wkt,Gopakumar:2016cpb} on the bootstrap program did not incorporate these contributions. In our context, such an ambiguity is generically associated with bulk contact interactions. We expect that other physical criteria such as consistency with OPE (see also~\cite{Gopakumar:2016cpb}) and locality of the dual bulk theory will constrain these contributions in a significant way. We do not go into this issue in detail, leaving it to future work.}~\cite{Polyakov:1974gs}
\begin{align}
\langle O_1(\bsk_1)O_2(\bsk_2)O_3(\bsk_3)O_4(\bsk_4)\rangle'
=\sum_n\left(W_n^{(\mathbf{s})}+W_n^{(\mathbf{t})}+W_n^{(\mathbf{u})}\right)\,,
\end{align}
where the label $n$ again runs over all the intermediate primary operators. The function $W_n^{(\mathbf{s})}$, which we call the $\mathbf{s}$-channel Polyakov block in the following, is a conformally covariant function enjoying the following two properties:
\begin{enumerate}
\item $W_n^{(\mathbf{s})}$ reproduces the non-analytic properties associated with the $\mathbf{s}$-channel factorization. More explicitly, we require
\begin{align}
\label{cond_disc}
\text{Disc}_{\mathbf{s}}W_n^{(\mathbf{s})}=T_{12;n}(\bsk_1,\bsk_2;-\bsk_{12})\,\text{Disc}_{\mathbf{s}}\langle O_n(\bsk_{12})O_n(-\bsk_{12})\rangle'T_{34;n}(\bsk_3,\bsk_4;-\bsk_{34})\,,
\end{align}
where we described the intermediate operators $O_n$ in the basis with diagonal two-point functions. Such a basis can easily be taken for scalar operators. Also, in Sec.~\ref{sec:spin}, we introduce the diagonal basis for spinning operators.

\item $W_n^{(\mathbf{s})}$ has no non-analyticity other than the one in the $\mathbf{s}$-channel collapsed limit $\mathbf{s}=0$. In particular, it is analytic in $\bsk_{13}$ and $\bsk_{14}$, so that it does not introduce $\mathbf{t},\mathbf{u}$-channel discontinuity.
\end{enumerate}
Also $W_n^{(\mathbf{t})}$ and $W_n^{(\mathbf{u})}$ are $\mathbf{t},\mathbf{u}$-channel analogues of $W_n^{(\mathbf{s})}$ and enjoy similar properties. In contrast to the ordinary conformal block, the crossing symmetry is manifest in this basis by construction. On the other hand, the consistency with the OPE is obscured, hence it gives a nontrivial constraint on the theory as Polyakov demonstrated~\cite{Polyakov:1974gs}.

\medskip
While Polyakov completed the bootstrap program in workable models elegantly, he did not derive an explicit form of the block $W_n^{(\mathbf{s})}$ because of technical complications due to conformal symmetry in momentum space, but rather he proposed a basis in position space.
Recently in~\cite{Gopakumar:2016wkt,Gopakumar:2016cpb}, Gopakumar et al. showed in Mellin space that the Polyakov block $W_n^{(\mathbf{s})}$ is nothing but the Witten exchange diagram. In the next subsection we revisit this equivalence and construct the Polyakov block in momentum space for the intermediate scalar case. Essentially because momentum space manifests the analyticity and factorization properties~\eqref{cond_disc}, our construction turns out to be quite simple and intuitive.

\subsection{Witten exchange diagram is Polyakov block}
\label{subsec:Witten}

We then construct the Polyakov block. Here we assume that the intermediate operator is not the identity. The identity operator case is discussed separately in the next subsection.

\paragraph{Lessons from a naive trial}

\begin{figure}[t]
\begin{center}
\includegraphics[width=50mm, bb=0 0 151 113]{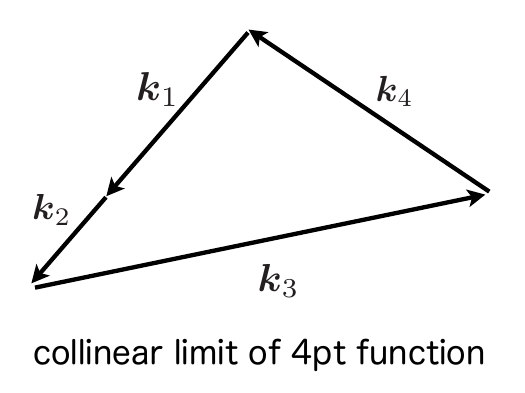}
\end{center}
\vspace{-5mm}
\caption{In the collinear limit, two external momenta are parallel and point the same direction.}
\label{fig:collinear_4pt}
\end{figure}

As a first step, it is instructive to start with a naive candidate for the Polyakov block of the form,
\begin{align}
\label{most_naive}
W_n^{(\mathbf{s})}\overset{?}{=}T_{12;n}(\bsk_1,\bsk_2;-\bsk_{12})\langle O_n(\bsk_{12})O_n(-\bsk_{12})\rangle'T_{34;n}(\bsk_3,\bsk_4;-\bsk_{34})\,,
\end{align}
which trivially satisfies the first requirement~\eqref{cond_disc}. However, it turns out to have undesired non-analyticity and does not satisfy the second criterion. As we mentioned in the previous section, the cubic vertex,
\begin{align}
&T_{12;3}(\bsk_1,\bsk_2;\bsk_3)
\nn\\*
&=-\frac{\Gamma(1-\nu_3)}{2^{\nu_3}}\frac{C_{123}}{C_{O_3O_3}}
\int_0^\infty \frac{dz}{z^{4}}\mathcal{B}_{\nu_1}(k_1;z)\mathcal{B}_{\nu_2}(k_2;z)z^{3/2}k_3^{-\nu_3}I_{\nu_3}(k_3z)\,,
\end{align}
is singular and non-analytic in the collinear limit $k_1+k_2=k_3$, essentially because of the exponential growth, $I_\nu(z)\sim e^z$ ($z\gg1$), of the Bessel $I$ function in the large $z$ region. This singularity leads to an undesired singularity of Eq.~\eqref{most_naive} in the collinear limit $k_1+k_2=k_{12}$ for example (see Fig.~\ref{fig:collinear_4pt}). The ansatz~\eqref{most_naive} then does not respect the second criterion. Our task is now to handle such an exponential growth of the Bessel $I$ function appropriately.

\paragraph{Witten exchange diagram}

We now proceed to showing that the improvement of the naive ansatz~\eqref{most_naive} is nothing but the Witten exchange diagram. To see this, let us first rewrite Eq.~\eqref{most_naive} as
\begin{align}
\nonumber
&\text{r.h.s. of } \eqref{most_naive}=\frac{\Gamma(1-\nu_n)\Gamma(1-\nu_n)}{2^{2\nu_n}}\frac{C_{12n}C_{34n}}{C_{O_nO_n}}\int_0^\infty \frac{dz_1}{z_1^{4}}\int_0^\infty \frac{dz_2}{z_2^{4}}
\\
\label{naive_integral}
&
\qquad\quad\,\,
\times
\mathcal{B}_{\nu_1}(k_1;z_1)\mathcal{B}_{\nu_2}(k_2;z_1)\mathcal{B}_{\nu_3}(k_3;z_2)\mathcal{B}_{\nu_4}(k_4;z_2)z_1^{3/2}z_2^{3/2}I_{\nu_n}(k_{12}z_1)I_{\nu_n}(k_{12}z_2)\,.
\end{align}
In the large $z_i$ region ($i=1,2$), the integrand behaves as
\begin{align}
\sim e^{-(k_1+k_2)z_1-(k_3+k_4)z_2+k_{12}(z_1+z_2)}\,,
\end{align}
which is the origin of undesired non-analyticities mentioned above. One possible way to remove this undesired exponential growth, without spoiling conformal covariance, is to replace the two Bessel $I$ functions by Bessel $K$ functions. This choice is essentially the ordinary conformal block. As we explain shortly, the ordinary conformal block has an additional $\mathbf{s}$-channel non-analyticity associated to the shadow operator, so that it does not satisfy the first criterion for the Polyakov block.

\medskip
The only thing we can do is then to replace one of the two Bessel $I$ functions by a Bessel $K$ function. For example, if we replace $I_\nu(k_{12}z_1)$ by $K_\nu(k_{12}z_1)$, the new integrand behaves in the large $z_i$ region as
\begin{align}
\sim e^{-(k_1+k_2)z_1-(k_3+k_4)z_2-k_{12}(z_1-z_2)}\,,
\end{align}
which is free from the undesired exponential growth when $z_1>z_2$. We therefore perform this replacement for the integral region $z_1>z_2$. Similarly, we replace $I_\nu(k_{12}z_2)$ by $K_\nu(k_{12}z_2)$ in the region $z_1<z_2$. Actually, this is nothing but what we usually do when constructing the bulk-to-bulk propagator. More explicitly, the bulk-to-bulk propagator in momentum space is given by\footnote{A standard normalization in the bulk is
\begin{align}
C_{OO}=-\frac{\Gamma(1-\nu)}{2^{2\nu-1}\Gamma(\nu)}\,,
\end{align}
which makes the prefactor in the first line to be $1$.}
\begin{align}
\cG_\nu(k;z_1,z_2)
&=
\frac{-\Gamma(1-\nu)}{C_{OO}\,2^{2\nu-1}\Gamma(\nu)}
\left[
\theta(z_1-z_2)z_1^{3/2}z_2^{3/2}K_\nu(kz_1)I_\nu(kz_2)+(1\leftrightarrow 2)
\right]
\\
\nonumber
&=\frac{\Gamma(1-\nu)\Gamma(1-\nu)}{C_{OO}\,2^{2\nu}}z_1^{3/2}z_2^{3/2}
\Big[I_\nu(kz_1)I_\nu(kz_2)
\\
\label{bulk-bulk-scalar}
&\qquad\qquad
-\theta(z_1-z_2)I_{-\nu}(kz_1)I_\nu(kz_2)
-\theta(z_2-z_1)I_{\nu}(kz_1)I_{-\nu}(kz_2)
\Big]\,,
\end{align}
where the second line exactly reproduces Eq.~\eqref{naive_integral} accompanied by the bulk-to-boundary propagators and integrals over the radial coordinates $z_i$. Note also that the third line is analytic in $k$. The Witten exchange diagram,
\begin{align}
\int_0^\infty \frac{dz_1}{z_1^{4}}\int_0^\infty \frac{dz_2}{z_2^{4}}
\mathcal{B}_{\nu_1}(k_1;z_1)\mathcal{B}_{\nu_2}(k_2;z_1)
\cG_{\nu_n}(k_{12};z_1,z_2)
\mathcal{B}_{\nu_3}(k_3;z_2)\mathcal{B}_{\nu_4}(k_4;z_2)\,,
\end{align}
therefore reproduces the $\mathbf{s}$-channel non-analyticity required by the factorization ansatz. Also, it has no other non-analyticity essentially because the bulk-to-bulk boundary propagator has an exponential suppression in the large $z_i$ region. Indeed, the $z$-ordered part of the propagator~\eqref{bulk-bulk-scalar} was originally added to remove the undesired exponential growth and thus undesired singularities of correlators. We therefore conclude that the Witten exchange diagram satisfies the two criteria of the Polyakov block and write
\begin{align}
W_n^{(\mathbf{s})}&=C_{12n}C_{34n}\int_0^\infty \frac{dz_1}{z_1^{4}}\int_0^\infty \frac{dz_2}{z_2^{4}}
\nn\\*
&\quad
\times
\mathcal{B}_{\nu_1}(k_1;z_1)\mathcal{B}_{\nu_2}(k_2;z_1)
\cG_{\nu_n}(k_{12};z_1,z_2)
\mathcal{B}_{\nu_3}(k_3;z_2)\mathcal{B}_{\nu_4}(k_4;z_2)\,.
\end{align}

\paragraph{Why not the conformal block?}

\begin{figure}[t]
\begin{center}
\includegraphics[width=140mm, bb=0 0 550 203]{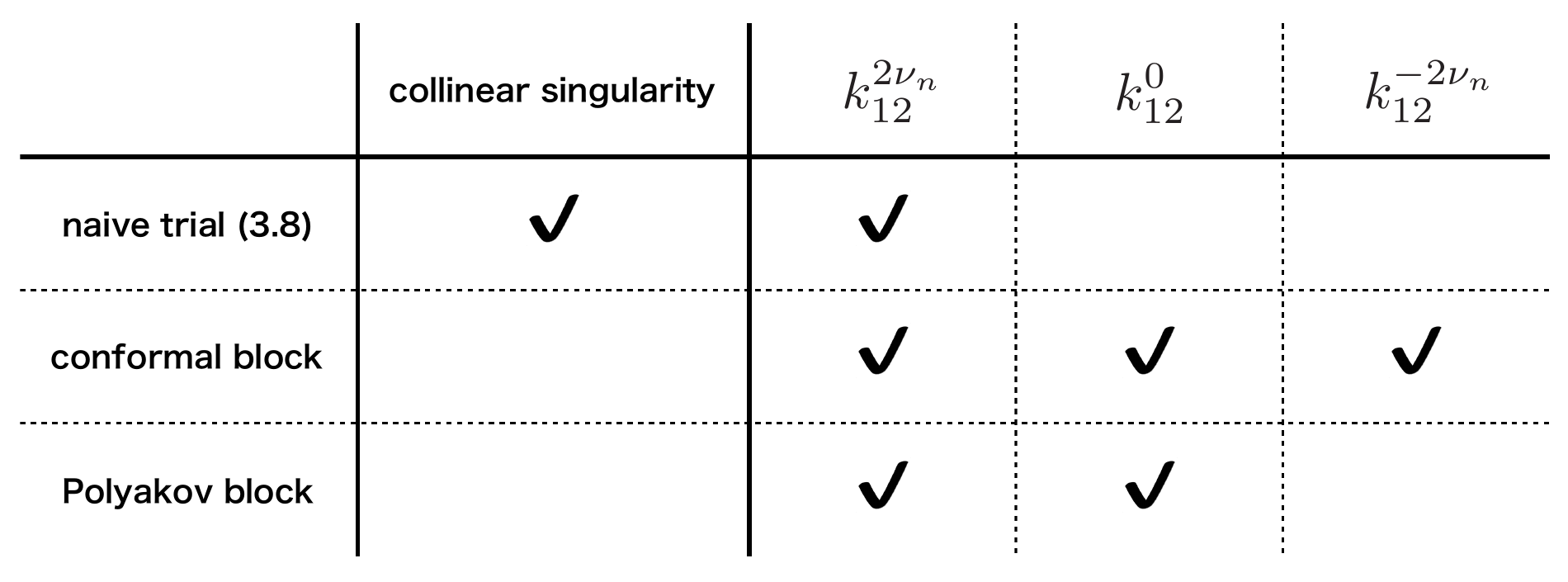}
\end{center}
\vspace{-5mm}
\caption{The naive trial~\eqref{most_naive} contains an undesired collinear singularity, even though its non-analytic property around $k_{12}=0$ is consistent with the Polyakov ansatz. On the other hand, both of the conformal block and the Polyakov block are free from the collinear singularity. These two are distinguished by the non-analyticity around $k_{12}=0$: The conformal block contains the non-analyticity $k_{12}^{-2\nu_n}$ associated with the shadow operator, whereas the Polyakov block does not.}
\label{fig:table}
\end{figure}

Before closing this subsection, let us make a brief remark on the difference of the Polyakov block and the ordinary conformal block. As we mentioned, if our goal were only to remove the undesired exponential growth in the large $z_i$ region, we could replace the two Bessel $I$ functions by Bessel $K$ functions, which gives the triple-$K$ integral squared. More precisely, conformal covariance requires the form,
\begin{align}
\label{ordinarly_block}
\frac{\langle O_1(\bsk_1)O_2(\bsk_2)O_n(-\bsk_{12})\rangle'\langle O_n(\bsk_{12})O_3(\bsk_3)O_4(\bsk_4)\rangle'}{\langle O_n(-\bsk_{12})O_n(\bsk_{12})\rangle'}\,,
\end{align}
which is regular except for the $\mathbf{s}$-channel collapsed limit $\bsk_{12}\to\boldsymbol{0}$. Since three-point functions in the squeezed limit scales as
\begin{align}
\langle O_1(\bsk_1)O_2(-\bsk_1)O_n(\boldsymbol{0})\rangle'\propto k_1^{\De_1+\De_2+\De_n-6}\,,
\end{align}
the collapsed limit $\bsk_{12}\to\boldsymbol{0}$ of Eq.~\eqref{ordinarly_block} is given by
\begin{align}
\eqref{ordinarly_block}\propto \frac{k_1^{\De_1+\De_2+\De_n-6}k_3^{\De_3+\De_4+\De_n-6}}{k_{12}^{2\nu_n}}
\quad
{\rm for}
\quad
k_{12}\to0
\,.
\end{align}
This non-analyticity is not the one required by the factorization ansatz $\text{Disc}_{s}W_n^{(\mathbf{s})} \propto k_{12}^{2\nu_n}$, but rather the one associated to the shadow operator $\propto k_{12}^{-2\nu_n}$ \cite{Ferrara:1972xe,Ferrara:1972ay,Ferrara:1972uq,Ferrara:1973vz,SimmonsDuffin:2012uy} (the shadow of $O_n$ has a scaling dimension $3-\Delta_n$, which corresponds to a sign flip $\nu_n\to-\nu_n$). Therefore, the ordinary conformal block does not satisfy the Polyakov ansatz. In this way, the absence of the non-analyticity associated with the shadow makes the Polyakov block different from the ordinary conformal block. See also Fig.~\ref{fig:table} for summary of the argument in this subsection.

\subsection{Polyakov blocks for an intermediate identity operator}
\label{subsec:identity}

Finally, let us consider the intermediate identity operator. The two-point function of the identity operator is $1$ in position space, so that the momentum space two-point function is
\begin{align}
\label{identity_2pt}
\langle 1(\bsk)\,1(-\bsk)\rangle'=(2\pi)^3\delta^{(3)}(\bsk)\,.
\end{align}
Here and in what follows, we denote the Fourier transform of the identity operator $1(\bsx)$ by $1(\bsk)$, even though it is simply $1(\bsk)=(2\pi)^3\delta^{(3)}(\bsk)$. Three-point functions involving the identity operator are also given by
\begin{align}
\langle O_1(\bsk_1)O_2(\bsk_2)\,1(\bsk_3)\rangle'=(2\pi)^3\delta^{(3)}(\bsk_3)C_{O_1O_2}k_1^{2\nu_1}=(2\pi)^3\delta^{(3)}(\bsk_3)\langle O_1(\bsk_1)O_2(\bsk_2)\rangle'\,,
\end{align}
where note that $C_{O_1O_2}$ may have a nonzero value only when $O_1$ and $O_2$ have the same scaling dimension. In the diagonal basis, $C_{O_1O_2}$ vanishes unless they are identical. Three-point functions have a delta function type non-analyticity in $\bsk_3$ and factorize as
\begin{align}
\langle O_1(\bsk_1)O_2(\bsk_2)\,1(\bsk_3)\rangle'=T_{12;{\rm id}}(\bsk_1,\bsk_2;\bsk_3)\langle 1(\bsk_3)\,1(-\bsk_3)\rangle'\,.
\end{align}
Here we introduced the cubic vertex $T_{12;{\rm id}}$ (``id" represents that the third operator is the identity operator) as
\begin{align}
T_{12;{\rm id}}(\bsk_1,\bsk_2;\bsk_3)=C_{O_1O_2}k_1^{2\nu_1}\,,
\end{align}
which is analytic in $\bsk_3$. The criteria in Sec.~\ref{subsec:Polyakov} then determine the $\mathbf{s}$-channel Polyakov block $W_{\rm id}^{(\mathbf{s})}$ with the intermediate identity operator as
\begin{align}
W_{\rm id}^{(\mathbf{s})}
&=T_{12;{\rm id}}(\bsk_1,\bsk_2;-\bsk_{12})\langle 1(\bsk_{12})\,1(-\bsk_{12})\rangle'
T_{34;{\rm id}}(\bsk_3,\bsk_4;-\bsk_{34})\,.
\end{align}

\medskip
Here it is worth mentioning that the Polyakov block for the intermediate identity operator is nothing but the disconnected part of four-point functions:
\begin{align}
W_{\rm id}^{(\mathbf{s})}
&=(2\pi)^3\delta^{(3)}(\bsk_{12})\langle O_1(\bsk_1)O_2(\bsk_2)\rangle'\langle O_3(\bsk_3)O_4(\bsk_4)\rangle'\,,
\end{align}
More explicitly, the Polyakov blocks with the intermediate identity operator reproduce the disconnected four-point functions as
\begin{align}
&\langle O_1(\bsk_1)O_2(\bsk_2)O_3(\bsk_3)O_4(\bsk_4)\rangle'|_{\rm disconnected}
\nn\\*
&=(2\pi)^3\delta^{(3)}(\bsk_{12})\langle O_1(\bsk_1)O_2(\bsk_2)\rangle'\langle O_3(\bsk_3)O_4(\bsk_4)\rangle'
+\text{2 terms}
\nn\\*
&=W_{\rm id}^{(\mathbf{s})}+W_{\rm id}^{(\mathbf{t})}+W_{\rm id}^{(\mathbf{u})}\,.
\end{align}
On the other hand, the connected correlators may be expanded as
\begin{align}
\langle O_1(\bsk_1)O_2(\bsk_2)O_3(\bsk_3)O_4(\bsk_4)\rangle'|_{\rm connected}
&=\sum_{n\neq \,{\rm id}}\left(W_n^{(\mathbf{s})}+W_n^{(\mathbf{t})}+W_n^{(\mathbf{u})}\right)
\,,
\end{align}
where the summation is over primary operators other than the identity operator.

\medskip
To contrast this property with the ordinary conformal block, let us consider the leading order of the so-called large $N$ CFT, whose four-point functions contain only the disconnected pieces as
\begin{align}
\label{4pt-largeN}
&\langle O_1(\bsk_1)O_2(\bsk_2)O_3(\bsk_3)O_4(\bsk_4)\rangle'
\nn\\*
&=(2\pi)^3\delta^{(3)}(\bsk_{12})\langle O_1(\bsk_1)O_2(\bsk_2)\rangle'\langle O_3(\bsk_3)O_4(\bsk_4)\rangle'
+\text{2 terms}
\,.
\end{align}
This four-point function can be expanded in, e.g., the $\mathbf{s}$-channel conformal blocks as
\begin{align}
&\langle O_1(\bsk_1)O_2(\bsk_2)O_3(\bsk_3)O_4(\bsk_4)\rangle'
\nn\\*
&=\sum_{n}\frac{\langle O_1(\bsk_1)O_2(\bsk_2)O_n(-\bsk_{12})\rangle'\langle O_n(\bsk_{12})O_3(\bsk_3)O_4(\bsk_4)\rangle'}{\langle O_n(\bsk_{12})O_n(-\bsk_{12})\rangle'}
\,,
\end{align}
where the intermediate operator $O_n$ runs over the double trace operators schematically of the form $\sim O\partial^nO$, in addition to the identity operator. Notice that the double trace operators originate from the decomposition of the $\mathbf{t}$- and $\mathbf{u}$-channel parts in \eqref{4pt-largeN}. On the other hand, in the crossing symmetric basis, we can expand the four-point functions as
\begin{align}
\langle O_1(\bsk_1)O_2(\bsk_2)O_3(\bsk_3)O_4(\bsk_4)\rangle'
&=W_{\rm id}^{(\mathbf{s})}+W_{\rm id}^{(\mathbf{t})}+W_{\rm id}^{(\mathbf{u})}\,.
\end{align}
Essentially because we are introducing the Polyakov blocks for all the channels, only the identity operator shows up as the intermediate operator.\footnote{
One might wonder why there are no contributions from the double trace operators in the crossing symmetric expansion of the large $N$ CFT. Here we briefly sketch how to see it from an explicit calculation (more details will be presented elsewhere~\cite{positivity}). In general, the double trace operator $O_{\rm d.t.}\sim O\partial^nO$ has a scaling dimension $2\Delta+n+\gamma$, where $\Delta$ is the dimension of $O$ and $\gamma$ is the anomalous dimension. In the large $N$ CFT, the anomalous dimension $\gamma$ vanishes just like free field theories. A careful evaluation of the triple-$K$ integral, the Polyakov block, and the ordinary conformal block then shows that
\begin{align}
&\lan O(\bsk_1) O(\bsk_2) O_{\text{d.t.}}(\bsk_3) \ran' =\widetilde{C}_{OOO_{\rm d.t.}}\mathcal{O}(\gamma^0)\,,
\quad
W_{O_{\rm d.t.}}^{(\mathbf{s})}=\widetilde{C}_{OOO_{\rm d.t.}}^2\mathcal{O}(\gamma^1)\,,
\nn\\*
&\frac{\langle O_1(\bsk_1)O_2(\bsk_2)O_{\rm d.t.}(-\bsk_{12})\rangle'\langle O_{\rm d.t.}(\bsk_{12})O_3(\bsk_3)O_4(\bsk_4)\rangle'}{\langle O_{\rm d.t.}(\bsk_{12})O_{\rm d.t.}(-\bsk_{12})\rangle'}=\widetilde{C}_{OOO_{\rm d.t.}}^2\mathcal{O}(\gamma^0)\,,
\end{align}
where we used the normalization factors $\widetilde{C}$ of position space and set $\widetilde{C}_{O_{\rm d.t.}O_{\rm d.t.}}=\mathcal{O}(1)$ for simplicity. Also, Eq.~\eqref{4pt-largeN} implies that $\widetilde{C}_{OOO_{\rm d.t.}}=\mathcal{O}(1)$. Hence, in the large $N$ CFT, i.e., in the limit $\gamma\to0$, the contributions from double trace operators vanish in the crossing symmetric expansion, whereas they do not in the conformal block expansion.}

\medskip
To summarize, in the crossing symmetric basis, the disconnected and connected four-point functions are manifestly separated into the Polyakov blocks for the identity and non-identity operators.
This property will be a big advantage for the study of large $N$ CFTs in particular. In large $N$ CFTs, the connected and disconnected correlators start at different orders in the $1/N$ expansion. As a result, in the ordinary conformal block expansion, the connected correlators are scrambled by the subleading corrections to the disconnected correlators. We expect that our new basis provides a complementary approach to large $N$ CFTs, which are holographically dual to weakly coupled bulk theories~\cite{positivity}.

\section{Extension to the case with an intermediate spinning operator}
\setcounter{equation}{0}
\label{sec:spin}

In this section we extend the construction of the crossing symmetric basis to the case with an intermediate spinning operator. The basic idea is the same as the intermediate scalar case, but there appears some technical complication due to the spin structure. To handle the spin structure, it is convenient to employ what we call the helicity basis, which is analogous to the helicity basis of massless particles. In Sec.~\ref{subsec:spin_2pt} we first summarize the basic properties of two-point functions in momentum space and introduce the helicity basis. In Sec.~\ref{subsec:spin_3pt} we provide a useful expression for three-point functions of two scalars and one tensor in the helicity basis. We then discuss their analytic properties and construct the crossing symmetric basis with an intermediate spinning operator in Sec.~\ref{subsec:spin_analytic}. Technical details of this section are collected in Appendices~\ref{App:2momentum} and~\ref{App-sec:3pt}.

\subsection{Two-point functions and helicity basis}
\label{subsec:spin_2pt}

A standard technique to handle symmetric traceless tensors in CFT is to contract all vector indices of the tensor operators with a null vector $\bsep$ called the polarization vector \cite{Bargmann:1977gy,Costa:2011mg}.\footnote{We use the bold and ordinary fonts for vectors and their components, e.g., $\bsep$ and $\ep^\mu$. Here $\mu$ is the vector index.} More explicitly, we introduce a shorthand notation \cite{Arkani-Hamed:2015bza},
\begin{align}
\ep^s.O=\ep^{\mu_1}\ep^{\mu_2}\ldots\ep^{\mu_s}O_{\mu_1\mu_2\ldots\mu_s}\,,
\end{align}
where $s$ is the spin of the operator $O$. In momentum space, two-point functions of primary operators with general spins are given by~\cite{Arkani-Hamed:2015bza} (see also Appendix~\ref{App-subsec:2pt} for the derivation)
\begin{align}
\label{2pt-mom}
\lan \ep^s.O(\bsk)\tilde{\ep}^s.O(-\bsk) \ran'=
{C}_{OO} k^{2\nu} \left( -\frac{(\bsep.\bsk)(\tilde\bsep.\bsk)}{k^2} \right)^s
P^{(\nu-s, -1/2)}_s\left( 1-\frac{k^2(\bsep.\tilde\bsep)}{(\bsep.\bsk)(\tilde\bsep.\bsk)} \right)\,,
\end{align}
where  $\De$ is the scaling dimension of $O$, two null vectors $\bsep$ and $\tilde{\bsep}$ are the polarization vectors, and $P^{(\alpha, \beta)}_n$ is the Jacobi polynomial. See Appendix~\ref{app:Jacobi} for the definition of the Jacobi polynomial and formulae used in this paper. Note that the corresponding two-point functions in position space take the form,
\begin{align}
\lan \ep^s.O(\bsx_1)\tilde{\ep}^s.O(\bsx_2) \ran 
= \widetilde{C}_{OO} \frac{[(\bsep.\tilde\bsep)x_{12}^2 - 2(\bsep.\bsx_{12})(\tilde\bsep.\bsx_{12})]^s}
{(x_{12}^2)^{\De+s}}\,,
\label{2pt-s-position}
\end{align}
where the normalization factor $\widetilde{C}_{OO}$ in position space is related to the momentum space one $C_{OO}$ by
\begin{align}
\label{Cpos-Cmom}
\widetilde{C}_{OO} = 2^{2\nu-s} \pi^{-3/2}  \frac{\Ga(\nu+s+3/2)}{s!\,\Ga(-\nu)} C_{OO}\,.
\end{align}
More details are in Appendix~\ref{App-subsec:2pt}, but the above is all what we need in the main text.

\medskip
When we work in momentum space, it is convenient to introduce what we call the helicity basis.
Let us parameterize the polarization null vectors as\footnote{Our definition of $\psi'$ is different from the one in~\cite{Arkani-Hamed:2015bza} by a sign factor: $\psi'_{\rm here}=-\psi'_{\rm there}$. Our definition of $\psi_{\rm here}'$ is associated with the rotation around $-\bsk$, whereas $\psi_{\rm there}'$ is associated with the rotation around $\bsk$.}
\begin{align}
\bsep=(\cos\psi,\sin\psi,i), \quad \tilde\bsep=(\cos\psi',-\sin\psi',-i)\,,
\label{nullhelicitychoice-3d}
\end{align}
where we set $\bsk=(0,0,k)$ without loss of generality by using the rotational invariance of the correlator. Angles $\psi$ and $\psi'$ represent rotation angles around the momenta $\bsk$ and $-\bsk$, respectively.
The $\bsep$-dependent factors in two-point functions are then reduced to the form,
\begin{align}
\frac{(\bsep.\bsk)(\tilde\bsep.\bsk)}{k^2}=1\,,
\quad
1-\frac{k^2(\bsep.\tilde\bsep)}{(\bsep.\bsk)(\tilde\bsep.\bsk)}=-\cos(\psi+\psi')\,,
\end{align}
so that the correlator \eqref{2pt-mom} may be reformulated as
\begin{align}
\lan \ep^s.O(\bsk)\tilde{\ep}^s.O(-\bsk) \ran'=
C_{OO} \left(k^2\right)^{\nu} (-)^s
P^{(\nu-s, -1/2)}_s\left( -\cos(\psi+\psi')\right)\,.
\end{align}
As we show in Appendix~\ref{hext-2pt}, we may rewrite it as~\cite{Dobrev:1977qv,Arkani-Hamed:2015bza}
\begin{align}
\lan \ep^s.O(\bsk)\tilde{\ep}^s.O(-\bsk) \ran'&= C_{OO} \left(k^2\right)^{\nu} 
\sum_{m=-s}^s {a_{s,\nu}(m)}\, e^{im(\psi+\psi')}\,,
\label{2ptmom-fourierbasis1}
\end{align}
where $m$ is identified with the rotation charge around the momentum of each operator, so that it is analogous to the helicity of massless particles. Based on this analogy, we call $m$ the helicity in the following. The coefficients $a_{s,\nu}(m)$ are defined by
\begin{align}
a_{s,\nu}(m)
 = \frac{\Ga(s+1/2) \Ga(s-\nu+1/2) \Ga(m+\nu+1/2)}{\sqrt{\pi} (s-m)! (s+m)!\,\Ga(\nu+1/2) \Ga(m-\nu+1/2)}
 \, ,
\end{align}
where $\displaystyle(z)_n=\frac{\Gamma(z+n)}{\Gamma(z)}$ is the shifted factorial (also known as the Pochhammer symbol).
Note  in particular that $a_{s,\nu}(m)=a_{s,\nu}(-m)$.
A detailed derivation of these results is provided in Appendix~\ref{App:2momentum}.

\medskip
Let us now identify each $m$ component in the summation~\eqref{2ptmom-fourierbasis1} with the contribution from the helicity $m$ operator. For this purpose, we define the helicity $m$ operator as
\begin{align}
O_m(\bsk)=\int_0^{2\pi}\frac{d\psi}{2\pi} e^{-im\psi} \ep^s.O(\bsk)\,,
\quad
O_m(-\bsk)=\int_0^{2\pi}\frac{d\psi'}{2\pi} e^{-im\psi'} \ep^s.O(\bsk)\,,
\end{align}
where the polarization vector $\bsep$ is defined through Eq.~\eqref{nullhelicitychoice-3d} upon an appropriate rotation. In this basis, the two-point function takes a simple form,
\begin{align}
\lan O_m(\bsk) O_{m'}(-\bsk) \ran'
&= \delta_{m,m'} C_{OO}  \,a_{s,\nu}(m) \left(k^2\right)^{\nu}\,.
\end{align}
Notice in particular that the helicity basis provides a diagonal basis for spinning operators.

\subsection{Three-point functions}
\label{subsec:spin_3pt}

We now proceed to three-point functions of two scalars and one symmetric traceless tensor. In position space, they are given by~\cite{Polyakov:1974gs}
\begin{align}
\label{3pt-s-position}
\lan O_1(\bsx_1)O_2(\bsx_2)\epsilon^s.O_3(\bsx_3) \ran 
= \widetilde{C}_{123} \frac{[x_{23}^2(\bsep.\bsx_{13}) - x_{13}^2(\bsep.\bsx_{23})]^s}
{(x_{12}^2)^{\frac{1}{2}(\De_{123}+s)}
(x_{13}^2)^{\frac{1}{2}(\De_{132}+s)}(x_{23}^2)^{\frac{1}{2}(\De_{231}+s)}}\,,
\end{align}
where $O_i$ ($i=1,2$) are primary scalars of dimensions $\De_i$, and $O_3$ is a spin $s$ primary symmetric traceless tensor of dimension $\De_3$. The momentum space correlators can be obtained either by Fourier transformations or by explicitly solving the conformal Ward-Takahashi identities in momentum space. In Appendix~\ref{App-sec:3pt}, we provide detailed derivations of them in both methods. Below, we summarize the results there.

\paragraph{Spin 1}
Before stating general results, it is instructive to elaborate the structure of three-point functions in the spin 1 and spin 2 cases. Since the denominator of Eq.~\eqref{3pt-s-position} takes a similar form to the scalar three-point functions, the three-point functions in momentum space will be expressed in terms of triple-$K$ type integrals. The polarization-dependent factor in the numerator will give derivatives in momenta. Indeed, for $s=1$, we find 
\begin{align}
&\lan O_1(\bsk_1)O_2(\bsk_2)\ep.O_3(\bsk_3) \ran'
\nn\\*
&
\propto
\left[
\bsep.(\bsk_1-\bsk_2)
+\bsep.(\bsk_1+\bsk_2)\frac{\mathfrak{D}_{123}}{\De_3-1}
\right]
\int_0^\infty \frac{dz}{z^{4}}z \, 
\mathcal{B}_{\nu_1}(k_1;z)\mathcal{B}_{\nu_2}(k_2;z)\mathcal{B}_{\nu_3}(k_3;z)\,,
\end{align}
where the last factor is a triple-$K$ type integral. Notice that its integrand has an additional factor~$z$ compared with scalar three-point functions. On the other hand, the prefactor inside the square brackets carries information of the polarization vector $\bsep$. As expected, this factor contains derivatives in the momenta $k_i$, which are packaged into a differential operator $\mathfrak{D}_{123}$ defined by
\begin{align}
\label{def_D_123}
\mathfrak{D}_{123}=\frac{k_1^2-k_2^2}{k_3^2}\left(
k_1\partial_{k_1}+k_2\partial_{k_2}
-\De_t+s+6
\right)
-\Big[(k_1\partial_{k_1}-\De_1)-(k_2\partial_{k_2}-\De_2)\Big]\,,
\end{align}
where the subscripts $123$ denote the labels of the operators $O_i$.
Note that when acting the operator $\mathfrak{D}_{123}$ on the triple-$K$ integral, we regard $k_1$, $k_2$ and $k_3$ as three independent variables. More explicitly, $\partial_{k_1}k_3=\partial_{k_2}k_3=0$ for example. Notice also that $\mathfrak{D}_{123}$ has an odd parity, $\mathfrak{D}_{123}\to-\mathfrak{D}_{123}$, under the exchange $1 \leftrightarrow 2$.

\paragraph{Spin 2}

Similarly, the spin 2 result is given by
\begin{align}
&\lan O_1(\bsk_1)O_2(\bsk_2)\ep^2.O_3(\bsk_3) \ran'
\nn\\*
&
\propto
\left[
(\bsep.\bsk_2)^2
+(\bsep.\bsk_2)(\bsep.\bsk_3)\frac{\De_3+\mathfrak{D}_{123}}{\De_3}
+(\bsep.\bsk_3)^2\frac{(\De_3+\mathfrak{D}_{123})(\De_3-2+\mathfrak{D}_{123})}{4\De_3(\De_3-1)}
\right]
\nn\\*
\label{spin2_momentum}
&\quad
\times\int_0^\infty \frac{dz}{z^{4}}z^2\mathcal{B}_{\nu_1}(k_1;z)\mathcal{B}_{\nu_2}(k_2;z)\mathcal{B}_{\nu_3}(k_3;z)\,,
\end{align}
where the integrand of the triple-$K$ type integral is accompanied by a factor $z^2$. The prefactor inside the square brackets is now a second order polynomial of the differential operator~$\mathfrak{D}_{123}$. Here one might wonder that $\bsk_1$ and $\bsk_2$ are not symmetric in the expression~\eqref{spin2_momentum}. However, it is easy to find that Eq.~\eqref{spin2_momentum} can be reformulated as
\begin{align}
&\lan O_1(\bsk_1)O_2(\bsk_2)\ep^2.O_3(\bsk_3) \ran'
\nn\\*
&
\propto
\left[
(\bsep.\bsk_1)^2
+(\bsep.\bsk_1)(\bsep.\bsk_3)\frac{\De_3-\mathfrak{D}_{123}}{\De_3}
+(\bsep.\bsk_3)^2\frac{(\De_3-\mathfrak{D}_{123})(\De_3-2-\mathfrak{D}_{123})}{4\De_3(\De_3-1)}
\right]
\nn\\*
\label{spin2_momentum'}
&\quad
\times\int_0^\infty \frac{dz}{z^{4}}z^2\mathcal{B}_{\nu_1}(k_1;z)\mathcal{B}_{\nu_2}(k_2;z)\mathcal{B}_{\nu_3}(k_3;z)\,.
\end{align}
Since $\mathfrak{D}_{123}$ has an odd parity under the exchange $1\leftrightarrow2$, it turns out that Eq.~\eqref{spin2_momentum} and equivalently Eq.~\eqref{spin2_momentum'} are consistent with the exchange symmetry of the two scalars.

\paragraph{General spin}

Three-point functions with a general spin $s$ accommodate similar structures as $s=1,2$ mentioned above. In Appendix~\ref{App-subsec:closed-form}, we derive a general expression, 
\begin{align}
\lan O_1(\bsk_1)O_2(\bsk_2)\ep^s.O_3(\bsk_3) \ran'
&=C_{123}\sum_{n=0}^s
\frac{s!}{n!(s-n)!}\left(\bsep.\bsk_2\right)^{s-n}(\bsep.\bsk_3)^{n}
\nn\\*
&\hspace{-8mm}
\times
\frac{\big(\tfrac{\De_3+s+\mathfrak{D}_{123}}{2}-n\big)_{n}}{(\De_3-1+s-n)_{n}}
\int_0^\infty \frac{dz}{z^{4}}z^s\mathcal{B}_{\nu_1}(k_1;z)\mathcal{B}_{\nu_2}(k_2;z)\mathcal{B}_{\nu_3}(k_3;z)\,,
\end{align}
where the triple-$K$ type integral has an additional $z^s$ factor. Also, the prefactor is an $s$-th order polynomial in $\mathfrak{D}_{123}$. Our normalization in momentum space is related to the position space one as
\begin{align}
\label{OPE_sst_mp_main}
C_{123}=\wt{C}{}_{123} \, 2^{1-s}\pi^3 i^s (\De_3-1)_s
\frac{\Gamma(\nu_1)\Gamma(\nu_2)\Gamma(\nu_3)}
{\Ga(\frac{\De_{123}+s}{2}) \Ga(\frac{\De_{231}+s}{2}) \Ga(\frac{\De_{312}+s}{2}) \Ga(\frac{\De_t+s-3}{2})}\,.
\end{align}
Just as we did in the spin $2$ case, we may rewrite it as
\begin{align}
\lan O_1(\bsk_1)O_2(\bsk_2)\ep^s.O_3(\bsk_3) \ran'
&=(-)^s \, C_{123}\sum_{n=0}^s
\frac{s!}{n!(s-n)!}\left(\bsep.\bsk_1\right)^{s-n}(\bsep.\bsk_3)^{n}
\nn\\*
&\hspace{-8mm}
\times
\frac{\big(\frac{\De_3+s-\mathfrak{D}_{123}}{2}-n\big)_{n}}{(\De_3-1+s-n)_{n}}
\int_0^\infty \frac{dz}{z^{4}}z^s\mathcal{B}_{\nu_1}(k_1;z)\mathcal{B}_{\nu_2}(k_2;z)\mathcal{B}_{\nu_3}(k_3;z)\,,
\end{align}
which is consistent with the exchange symmetry of the two scalars. In particular we notice that three-point functions vanish when two scalars are identical and the tensor $O_3$ has an odd spin, as is implied also by the position space result~\eqref{3pt-s-position}.

\paragraph{Three-point functions in helicity basis}

Finally, let us introduce three-point functions in the helicity basis. Without loss of generality, we set the third momentum $\bsk_3$ and the polarization vector $\bsep$ as
\begin{align}
\label{k3ep}
\bsk_3=(0,0,k_3)\,,
\quad
\bsep=(\cos\psi,\sin\psi,i)\,.
\end{align}
It is convenient to parametrize the momentum $\bsk_2$ as
\begin{align}
\label{k2}
\bsk_2 = k_2(\cos\chi\sin\te, \sin\chi\sin\te, \cos\te)\,.
\end{align}
The polarization-dependent factor is then written as
\begin{align}
\bsep.\bsk_2 = k_2 \left( \cos(\psi-\chi)\sin\te+i\cos\te \right)
\,,
\quad
\bsep.\bsk_3=ik_3\,.
\label{epk2-epk3}
\end{align}
Therefore, three-point functions involving the helicity $m$ tensor, $O_{3_m}$, are (see also Fig.~\ref{fig:3pt_Witten_helicity})
\begin{align}
&\lan O_1(\bsk_1)O_2(\bsk_2)O_{3_m}(\bsk_3) \ran'
\nn\\*
&~~ =\int_0^{2\pi} \frac{d\psi}{2\pi} \, e^{-im\psi} \lan O_1(\bsk_1)O_2(\bsk_2)\ep^s.O_3(\bsk_3) \ran'
\nn\\*
\label{00s_helicity}
&~~ =  \cA_{123_m}(\bsk_1,\bsk_2,\bsk_3;\mathfrak{D}_{123}) 
\int_0^\infty \frac{dz}{z^{4}}z^s\mathcal{B}_{\nu_1}(k_1;z)\mathcal{B}_{\nu_2}(k_2;z)\mathcal{B}_{\nu_3}(k_3;z) \,.
\end{align}
Here we introduced a differential operator $\cA_{123_m}$ for $-s \leq m \leq s$ as (see Appendix~\ref{app:A123m} for derivation)
\begin{align}
&
\cA_{123_m}(\bsk_1,\bsk_2,\bsk_3;\mathfrak{D}_{123})
= C_{123} \frac{i^{s-|m|} s!}{2^{|m|} |m|! (s-|m|)!} \nn\\
& \qquad \times
 e^{-im\chi}\sum_{n=0}^{s-|m|} \frac{(s-|m|)!}{n!(s-|m|-n)!} k_2^{s-n} \wh{P}_{s-n,|m|}(\cos\te) \, k_3^n 
\frac{\big(\tfrac{\De_3+s+\mathfrak{D}_{123}}{2}-n\big)_{n}}{(\De_3-1+s-n)_{n}} \,,
\label{A_123m}
\end{align}
where $\wh{P}_{\ell,|m|}(\cos\te)$ is proportional to the associated Legendre function and normalized as Eq.~\eqref{def:Phat}.
In the next subsection, we clarify analytic properties of three-point functions involving one tensor based on the expression~\eqref{00s_helicity}. As the appearance of triple-$K$ type integrals implies, we find that the argument goes quite similar to the scalar case discussed in the previous section.

\subsection{Construction of crossing symmetric basis}
\label{subsec:spin_analytic}

We proceed to discussing analytic properties of three-point functions~\eqref{00s_helicity} and constructing the crossing symmetric basis. 
First, according to the parametrizations \eqref{k3ep} and \eqref{k2}, the functions $k_2^{s-n} \wh{P}_{s-n,|m|}(\cos\te)$ is a polynomial in $\bsk_2$. Next, the differential operator $\mathfrak{D}_{123}$ enters Eq.~\eqref{A_123m} in the combination $k_3\mathfrak{D}_{123}$,
which is analytic because $(k_1^2-k_2^2)/k_3=-(\bsk_1-\bsk_2).\bsk_3/k_3=-k_{1z}+k_{2z}$ (see Eq.~\eqref{def_D_123}).
We then conclude that the operator $\cA_{123_m}(\bsk_1,\bsk_2,\bsk_3;\mathfrak{D}_{123})$ is a polynomial in the momenta $\bsk_i$ and the Euler operators $k_i\partial_{k_i}$. Since the Euler operator does not introduce new non-analyticity, non-analytic properties of three-point functions are essentially captured by the triple-$K$ integral.

\paragraph{Cubic vertex}

Let us then determine the cubic vertex. Just as the scalar three-point function case, our starting point is the relation,
\begin{align}
&{\rm Disc}_{k_3^2}\int_0^\infty \frac{dz}{z^{4}}z^s\mathcal{B}_{\nu_1}(k_1;z)\mathcal{B}_{\nu_2}(k_2;z)\mathcal{B}_{\nu_3}(k_3;z)
\nn\\*
\label{triple-K_n.a.}
&=-\frac{\Gamma(1-\nu_3)}{2^{\nu_3}}\int_0^\infty \frac{dz}{z^{4}}z^s\mathcal{B}_{\nu_1}(k_1;z)\mathcal{B}_{\nu_2}(k_2;z)z^{3/2}k_3^{-\nu_3}I_{\nu_3}(k_3z)\times{\rm Disc}_{k_3^2}\left(k_3^2\right)^{\nu_3}\,.
\end{align}
Notice again that the $I_{\nu_3}$ part of the bulk-to-boundary propagator is responsible for the non-analytic properties around $k_3=0$, whereas the $I_{-\nu_3}$ part is for removing the singularity in the collinear limit $k_1+k_2=k_3$. As we mentioned, the differential operator $\mathcal{A}_{123_m}$ does not produce any new non-analyticity, so that the non-analytic parts of three-point functions are obtained by acting $\mathcal{A}_{123_m}$ on the r.h.s. of Eq.~\eqref{triple-K_n.a.}. Moreover, the differential operator $\mathfrak{D}_{123}$ does not contain derivatives in $k_3$. Therefore, the operator $\mathcal{A}_{123_m}$ does not change non-analytic properties around $k_3=0$. All in all, we arrive at the factorization relation,
\begin{align}
\text{Disc}_{k_3^2}\langle O_1(\bsk_1)O_2(\bsk_2)O_{3_m}(\bsk_3)\rangle'
&=T_{12;3_m}(\bsk_1,\bsk_2;\bsk_3)\text{Disc}_{k_3^2}\langle O_{3_{m}}(-\bsk_3)O_{3_m}(\bsk_3)\rangle'
\,,
\end{align}
where the cubic vertex $T_{12;3_m}$ is given by
\begin{align}
&T_{12;3_m}(\bsk_1,\bsk_2;\bsk_3)
\nn\\*
&=-\frac{\Gamma(1\!-\!\nu_3)}{2^{\nu_3}}\frac{\cA_{123_m}(\bsk_1,\bsk_2,\bsk_3;\mathfrak{D}_{123})}{C_{O_3O_3}a_{s,\nu_3}(m)}
\int_0^\infty \frac{dz}{z^{4}}\mathcal{B}_{\nu_1}(k_1;z)\mathcal{B}_{\nu_2}(k_2;z)z^{3/2}k_3^{-\nu_3}I_{\nu_3}(k_3z)\,.
\label{cubic_V_spin}
\end{align}
Notice that the cubic vertex is analytic at $k_3=0$, just as the scalar three-point case.
We can also show that the cubic vertex satisfies the conformal WT identities for $O_1, O_2$ and the shadow of $O_3$ by applying the argument given in Appendix~\ref{App-subsec:WTa0}.

\paragraph{Crossing symmetric basis}

Finally, let us construct the crossing symmetric basis. In the helicity basis, two-point functions are diagonal with respect to the spin and helicity, so that the requirement~\eqref{cond_disc} for the $\mathbf{s}$-channel Polyakov block is simply carried over as
\begin{align}
&\text{Disc}_{\mathbf{s}}W_{n_m}^{(\mathbf{s})}
\nn\\*
&=T_{12;n_m}(\bsk_1,\bsk_2;-\bsk_{12})\,\text{Disc}_{\mathbf{s}}\langle O_{n_m}(\bsk_{12})O_{n_m}(-\bsk_{12})\rangle'T_{34;n_m}(\bsk_3,\bsk_4;\bsk_{12})
\label{cond_disc_helicity}
\,,
\end{align}
where $W_{n_m}^{(\mathbf{s})}$ is the Polyakov block with an intermediate operator $O_{n_m}$ with spin $s$, helicity $m$, and scaling dimension $\Delta_n$. $T_{12;n_m}$ and $T_{34;n_m}$ are the cubic vertices we have just introduced above. As we mentioned in Sec.~\ref{subsec:Polyakov}, we also require that the $\mathbf{s}$-channel block has no other non-analyticity. Since $\mathcal{A}_{12n_m}$ and $\mathcal{A}_{34n_m}$ in the cubic vertices do not change the non-analytic properties, we may apply the argument in Sec.~\ref{subsec:Witten} in a straightforward manner to conclude
\begin{align}
&W_{n_m}^{(\mathbf{s})}=
\frac{\cA_{12n_m}(\bsk_1,\bsk_2,-\bsk_{12};\mathfrak{D}_{12n})\cA_{34n_m}(\bsk_3,\bsk_4,\bsk_{12};\mathfrak{D}_{34n})}{a_{s,\nu_n}(m)}
\nn\\*
&\quad
\times
\int_0^\infty
\frac{dz_1}{z_1^{4-s}}
\int_0^\infty
\frac{dz_2}{z_2^{4-s}}
\mathcal{B}_{\nu_1}(k_1;z_1)\mathcal{B}_{\nu_2}(k_2;z_1)
\cG_{\nu_n}(k_{12};z_1,z_2)
\mathcal{B}_{\nu_3}(k_3;z_2)\mathcal{B}_{\nu_4}(k_4;z_2)\,,
\label{Polyakov_helicity}
\end{align}
where the second line is the Polyakov block for an intermediate scalar with an additional factor $z_1^sz_2^s$ in the integrand. $\cG_{\nu_n}(k_{12};z_1,z_2)$ is the {\it scalar} bulk-to-bulk propagator defined by Eq.~\eqref{bulk-bulk-scalar}. On top of it, we have the helicity dependent differential operators $\cA_{12n_m}$ and $\cA_{34n_m}$ (see also Fig.~\ref{fig:Polyakov_helicity}).
It is easy to see that Eq.~\eqref{Polyakov_helicity} enjoys the $\mathbf{s}$-channel factorization property~\eqref{cond_disc_helicity} and does not have any other non-analyticities. The $\mathbf{t}$- and $\mathbf{u}$-channel Polyakov blocks are also be defined in a similar fashion.

\section{Conclusion}
\setcounter{equation}{0}

In this paper we explicitly constructed the crossing symmetric basis for conformal four-point functions of primary scalars, following Polyakov's analyticity based ansatz:
\begin{align}
\langle O_1(\bsk_1)O_2(\bsk_2)O_3(\bsk_3)O_4(\bsk_4)\rangle'
=\sum_n\left(W_n^{(\mathbf{s})}+W_n^{(\mathbf{t})}+W_n^{(\mathbf{u})}\right)\,.
\end{align}
By requiring consistent factorization in each channel, we showed that the Polyakov block $W_n^{(\mathbf{s})}$ for an intermediate scalar is nothing but the Witten exchange diagram. The Polyakov block for an intermediate spinning operator is given by Eq.~\eqref{Polyakov_helicity}, as a natural extension of the intermediate scalar case.
As our construction demonstrated, momentum space provides a natural language for this program, essentially because analytic structures are manifest in momentum space.
On the way to construction, we also found the new closed expressions for momentum space three-point functions of two scalars and one tensor.
Thanks to these expressions, the Polyakov blocks for a spinning intermediate operator can be obtained by simply acting the differential operator Eq.~\eqref{A_123m} on the Witten exchange diagram for an intermediate scalar, as depicted in Fig.~\ref{fig:Polyakov_helicity}. This is our main result.

\medskip
As a concluding remark, we would like to present several promising future directions. First, we have focused on three dimensional space in the present paper, for technical simplicity. Even though there exist technical complications associated to spins and helicities, it is conceptually straightforward to extend our argument to higher dimensions. We will present this result in a forthcoming paper~\cite{crossing2}. It will also be interesting to construct the crossing symmetric basis for conformal correlators of (external) spinning operators such as the energy-momentum tensor.
These explicit forms of the crossing symmetric basis will be useful to revisit and generalize Polyakov's original bootstrap approach.
Besides, as we discussed in Sec.~\ref{subsec:identity}, the connected and disconnected correlators are manifestly separated in our basis, in contrast to the ordinary conformal block. We believe that this is a big advantage when discussing the large $N$ CFTs, which are holographically dual to weakly coupled bulk theories. We will use this property to revisit positivity bounds on effective interactions in the bulk~\cite{positivity}. Another interesting direction will be application to cosmology. Recent progress in cosmology has shown that (non-)analytic structures of higher-point correlators are useful to probe new particles coupled to the inflaton sector, as is dubbed the ``cosmological collider physics" program~\cite{Chen:2009zp, Baumann:2011nk, Noumi:2012vr,Arkani-Hamed:2015bza}. Indeed, some of the momentum space techniques used in this paper were developed in the context of cosmic inflation.
It will be useful to introduce a basis for inflationary correlators by extending our construction. We hope to revisit these issues in near future.

\appendix

\subsection*{Note added}

While this paper was being finalized, Ref.~\cite{Costa:2018mcg} appeared on arXiv. There, what they called the weight shifting operator was introduced in position space, whose action on scalar Witten exchange diagrams gives those involving spinning fields (see also~\cite{Sleight:2016hyl,Sleight:2017fpc,Isono:2017grm} for other differential operators in position space with a similar property). This seems similar to the property of our differential operator $\mathcal{A}$ defined in momentum space as Eq.~\eqref{00s_helicity}, even though these two operators have different origins and their relation is not clear. It would be interesting to explore a possible connection thereof.

\section*{Acknowledgments}
HI would like to thank Yuki Sato for fruitful discussions. TN would like to thank Masatoshi Noumi for valuable discussions on Appell's hypergeometric function $F_4$ and related topics. HI is supported in part by the ``CUniverse'' research promotion project by Chulalongkorn University (grant reference CUAASC), in part by Franco-Thai Cooperation Program in Higher Education and Research `PHE SIAM 2016'.
We also thank the Yukawa Institute for Theoretical
Physics at Kyoto University for hospitality during the workshop YITP-W-17-08 on ``Strings and Fields 2017," where a part of this work was done.
TN  is in part supported by JSPS KAKENHI Grant Numbers JP17H02894 and JP18K13539, and MEXT KAKENHI Grant Number JP18H04352.
GS is supported in part by the DOE grant DE-SC0017647 and the Kellett Award of the University of Wisconsin.

\section{Two-point functions with spins in momentum space}
\label{App:2momentum}
\setcounter{equation}{0}

In this and the next appendices, we derive two- and three-point functions with spins in momentum space.
For generality, we work in general dimension $d=2h$ in these appendices unless otherwise stated, though we set $d=3$ in the main part.

\subsection{Two-point functions by Fourier transformation}
\label{App-subsec:2pt}

Let us compute the two-point function $\lan \ep^s.O(\bsk)\tilde{\ep}^s.O(-\bsk) \ran'$ in this appendix.
This has been computed in \cite{Dobrev:1977qv} by Fourier transformation in general dimensions and in \cite{Arkani-Hamed:2015bza} by solving the Ward-Takahashi (WT) identities in three dimension. Here we review the derivation by Fourier transformation, which enables us to fix the relative overall constant between position-space and momentum-space correlators.

\medskip
Our convention for Fourier transformation is
\begin{align}
\label{OkOx}
O(\bsk)=\int{d^dx} \, e^{-i\bsk\cdot\bsx} O(\bsx)\,,
\end{align}
where $O(\bsx)$ is the position space operator and $O(\bsk)$ is the momentum space one.
Combining this with the two-point function in position space \eqref{2pt-s-position} gives 
\begin{align}
\label{2pt-mom-1}
&{}
{\lan \ep^s.O(\bsk)\tilde{\ep}^s.O(-\bsk) \ran'} \nn\\
&{} \quad
= \widetilde{C}_{OO} \sum_{a=0}^s
\binom{s}{a}
\frac{2^{-2\nu-a} \pi^h \,\Ga(-\nu-a)}{\Ga(\nu+h+a)} (\bsep.\tilde\bsep)^{s-a} 
\left(\bsep.\frac{\pd}{\pd\bsk}\right)^a \left(\tilde\bsep.\frac{\pd}{\pd\bsk}\right)^a (k^2)^{\nu+a}\,.
\end{align}
Here we applied the binomial expansion after replacing position coordinates in the numerator of Eq.~\eqref{2pt-s-position} by momentum derivatives,
and then used the Fourier transformation of $(x^2)^{-\al}$ given by the formula,
\begin{align}
\label{x2power-FT}
\int d^dx \, e^{-i\bsk.\bsx} (x^2)^{-\al} = 2^{2(h-\al)} \pi^h \frac{\Ga(h-\al)}{\Ga(\al)} (k^2)^{\al-h}\,.
\end{align}
We also defined $\nu=\De-h$. The momentum derivatives can be rewritten as $(c=-\nu-a)$
\begin{align}
\hspace{-1mm}\left(\bsep.\frac{\pd}{\pd\bsk}\right)^a \left(\tilde\bsep.\frac{\pd}{\pd\bsk}\right)^a \frac{1}{(k^2)^{c}}
&= \left(\bsep.\frac{\pd}{\pd\bsk}\right)^a \left[
(-2)^{a}\frac{\Gamma(c+a)}{\Gamma(c)}
\frac{ (\tilde\bsep.\bsk)^a}{(k^2)^{c+a}}
\right]
\nn\\*
&= \sum_{m=0}^a \frac{(-2)^{a+m} (a!)^2 \, \Ga(c+a+m)}{(m!)^2(a-m)! \, \Ga(c)}
\frac{(\bsep.\tilde\bsep)^{a-m} (\bsep.\bsk)^m (\tilde\bsep.\bsk)^m}{(k^2)^{c+a+m}}\,,
\end{align}
where we used $\bsep.\bsep=\tilde{\bsep}.\tilde{\bsep}=0$.
Applying this and the resummation $\displaystyle\sum_{a=0}^s \sum_{m=0}^a = \sum_{m=0}^s \sum_{a=m}^s$ to Eq.~\eqref{2pt-mom-1} gives
\begin{align}
\label{2pt-mom-2}
&{}
\lan \ep^s.O(\bsk)\tilde{\ep}^s.O(-\bsk) \ran'
= \widetilde{C}_{OO} \, (k^2)^\nu 
\nn\\
&{} \quad
\times\sum_{m=0}^s
\binom{s}{m}
\frac{2^{m-2\nu}\pi^h\Ga(m-\nu)\Ga(h+\nu+s-m-1)}{\Ga(h+\nu-1)\Ga(h+\nu+s)}
(\bsep\cd\tilde\bsep)^{s-m} 
\left[ \frac{(\bsep.\bsk) (\tilde\bsep.\bsk)}{k^2} \right]^m\,,
\end{align}
where we used the following summation formula
\begin{align}
\sum_{a=m}^s \frac{a!}{(s-a)!(a-m)!} \frac{(-1)^a}{\Ga(\nu+h+a)}
= \frac{(-1)^m m! \, \Ga(h+\nu+s-m-1)}{(s-m)! \, \Ga(h+\nu-1)\Ga(h+\nu+s)}\,.
\end{align}
Changing the summation index $m \to s-m$, using the definition of the Jacobi polynomial \eqref{jacobi-def}, and applying the formula $\Ga(-\nu)\Ga(\nu+1)=-\pi/\sin(\pi\nu)$, we arrive at
\begin{align}
\lan \ep^s.O(\bsk)\tilde{\ep}^s.O(-\bsk) \ran'=
{C}_{OO} \, (k^2)^\nu  \left( -\frac{(\bsep.\bsk)(\tilde\bsep.\bsk)}{k^2} \right)^s
P^{(\nu-s, h-2)}_s\left( 1-\frac{k^2(\bsep.\tilde\bsep)}{(\bsep.\bsk)(\tilde\bsep.\bsk)} \right)\,, 
\end{align}
where the overall constant $C_{OO}$ is given by
\begin{align}
C_{OO}=2^{s-2\nu}\pi^h\frac{s!\,\Gamma(-\nu)}{\Gamma(\nu+s+h)}\widetilde{C}_{OO}\,.
\end{align}
This result for $h=3/2$ gives Eq.~\eqref{2pt-mom} in the main text.

\subsection{Helicity basis}
\label{hext-2pt}
The expression \eqref{2pt-mom} is compact, but to understand the helicity dependence, it is more useful to rewrite it in such a form that the angles among momenta and helicity vectors are manifest~\cite{Dobrev:1977qv,Arkani-Hamed:2015bza}.
In this subsection we concentrate only on the three-dimensional case. A general dimensional version will be presented in a forthcoming paper \cite{crossing2}. 

\medskip
To simplify the analysis, we set the null helicity vectors to \eqref{nullhelicitychoice-3d} by using the {\rm O(3)}-rotational invariance. 
The two-point function \eqref{2pt-mom} then becomes
\begin{align}
\label{2ptmom-AHMbasis}
C_{OO} (k^2)^\nu (-)^{s}
P_s^{(\nu-s,-1/2)}(-\cos(\psi+\psi'))\,.
\end{align}
We can easily see that the Jacobi polynomial is defined on the unit circle and only depends on the angle $\psi+\psi'$. 
Furthermore, $P^{(\nu-s,-1/2)}(t)$ is a polynomial of degree $s$, and any power $\cos^m(\psi+\psi')$ can be expanded by $\{\cos n(\psi+\psi')\}$ with $n=0,1,\cdots,m$. Therefore, the Jacobi polynomial $P_s^{(\nu-s,-1/2)}(-\cos(\psi+\psi'))$ can be expanded as
\begin{align}
P_s^{(\nu-s,-1/2)}(-\cos(\psi+\psi')) = {\alpha}_0 + 2\sum_{m=1}^s {\alpha}_m \cos m(\psi+\psi'),
\end{align}
where the coefficients ${\alpha}_m$ are given by
\begin{align}
\label{am-integral}
{\alpha}_m  &= \int_0^{2\pi} \frac{d\te}{2\pi} (\cos m\te) P_s^{(\nu-s,-1/2)}(-\cos\te) 
\nn\\
&
= (-)^s\frac{\Ga(s+1/2)\Ga(s-\nu+1/2) \Ga(m+\nu+1/2)}{\sqrt{\pi} (s-m)! (s+m)!\,\Ga(\nu+1/2) \Ga(m-\nu+1/2)}\,.
\end{align}
The integration \eqref{am-integral} was carried out using the integration formula \eqref{jacobi-int} after rewriting $\cos m\te$ as a Chebyshev polynomial defined in Eq.~\eqref{Chebyshev}.
Substituting these results into the two-point function \eqref{2ptmom-AHMbasis}, we find
\begin{align}
\lan \ep^s.O(\bsk) \tilde\ep^sO(-\bsk) \ran'
&= C_{OO} (k^2)^\nu
\sum_{m=-s}^s a_{s,\nu}(m) e^{im(\psi_1+\psi_2)}\,,
\label{2ptmom-fourierbasis2}
\end{align}
where we introduced $a_{s,\nu}(m)=(-)^s{\alpha}_m$ for $m\geq0$ and $a_{s,\nu}(m)=a_{s,\nu}(-m)$ for $m<0$.

\section{Three-point functions of two scalars and one tensor}
\label{App-sec:3pt}

In this appendix we derive two types of closed forms for three-point functions involving two scalars and one tensor. The first expression is the one we used in the main text:\footnote{As we mentioned in the previous section, we work in general dimension $d=2h$ in this section.}
\begin{align}
\lan O_1(\bsk_1)O_2(\bsk_2)\ep^s.O_3(\bsk_3) \ran'
&=C_{123}\sum_{n=0}^s
\frac{s!}{n!(s-n)!}\left(\bsep.\bsk_2\right)^{s-n}(\bsep.\bsk_3)^{n}
\nn\\*
\label{3pt_spin_generealD}
&\hspace{-10mm}
\times\frac{\big(\tfrac{\De_3+s+\mathfrak{D}_{123}}{2}-n\big)_{n}}{(\De_3-1+s-n)_{n}}\int_0^\infty \frac{dz}{z^{d+1}}z^s\mathcal{B}_{\nu_1}(k_1;z)\mathcal{B}_{\nu_2}(k_2;z)\mathcal{B}_{\nu_3}(k_3;z)\,,
\end{align}
where $\mathfrak{D}_{123}$ is a differential operator in $k_i$'s, defined by Eq.~\eqref{fkD_generalD}. As we discussed in Sec.~\ref{sec:spin}, this expression manifests analytic properties of three-point functions, so that it is useful for our construction of the crossing symmetric basis. We also derive the other form,
\begin{align}
\nonumber
&\lan O_1(\bsk_1)O_2(\bsk_2)\ep^s.O_3(\bsk_3) \ran'
\\
& \quad
= c_{123} 
\sum_{m_1+m_2=s} \sum_{q_1=0}^{m_1} \sum_{q_2=0}^{m_2}
d_{123}(q_1,q_2,m_1,m_2)\,(-)^{m_2}
\nn\\
\label{3pt_spin_generealD_FT}
&\qquad\qquad\times
\, (\bsep.\bsk_1)^{m_1-q_1} (\bsep.\bsk_2)^{m_2-q_2} \,(-\bsep.\bsk_3)^{q_1+q_2}
\cI_{s}(\nu_1+q_1, \, \nu_2+q_2, \, \nu_3-q_1-q_2)\,,
\end{align}
where $c_{123}$ and $d_{123}$ are numerical coefficients given by Eq.~\eqref{c123} and Eq.~\eqref{A123}, respectively, and $\cI_{s}$ is a triple-$K$ type integral,
\begin{align}
\cI_s(\al_1,\al_2,\al_3)
= \int_0^\infty \frac{dz}{z^{d+1}}z^s\mathcal{B}_{\al_1}(k_1;z)\mathcal{B}_{\al_2}(k_2;z)\mathcal{B}_{\al_3}(k_3;z)\,.
\end{align}
As we discuss in Appendix~\ref{App-subsec:3pt-FT}, this expression naturally arises when we perform Fourier transformation of the position space correlators. Even though the derivation itself is a bit technical, the two closed forms are a part of the main results of this paper. The rest of this appendix goes along the line of the derivation of the first expression~\eqref{3pt_spin_generealD}.

\medskip
Following the argument in Appendix C of \cite{Arkani-Hamed:2015bza}, we begin with a general ansatz,\footnote{
Our definition of $a_n$ is slightly different from the one in~\cite{Arkani-Hamed:2015bza}: $a_{n}^{(\rm here)}=a_{s-n}^{(\rm there)}$.}
\begin{align}
\label{00s_ansatz}
\lan O_1(\bsk_1)O_2(\bsk_2)\ep^s.O_3(\bsk_3) \ran'
&=k_3^{\Delta_t-2d-s}\sum_{n=0}^s\gamma^{s-n}\delta^n a_n(\kappa_1,\kappa_2)\,,
\end{align}
where $\De_t=\De_1+\De_2+\De_3$ and we introduced
\begin{align}
\gamma=\bsep.\bsk_1-\bsep.\bsk_2\,,
\quad
\delta=\bsep.\bsk_1+\bsep.\bsk_2\,,
\quad
\kappa_i=\frac{k_i}{k_3}\quad
(i=1,2)\,.
\end{align}
Notice that we already used the dilatation symmetry to fix the exponent of $k_3$. Also it is easy to show that Eq.~\eqref{00s_ansatz} provides the most general ansatz consistent with the dilatation symmetry and the polarization vector dependence. Our task is now to determine the functions $a_n(\kappa_1,\kappa_2)$. In Appendix~\ref{App-subsec:3pt-rec}, we first derive a recursion relation for $a_n$ by using a subset of special conformal Ward-Takahashi (WT) identities. We then solve the recursion relation explicitly in Appendix~\ref{App-subsec:closed-form}. The initial condition for this program will be given in Appendix~\ref{App-subsec:WTa0}. The relation between the momentum space correlator and the position space one is discussed in Appendix~\ref{App-subsec:3pt-FT} by performing Fourier transformation.

\subsection{Recursion relations for $a_n$}
\label{App-subsec:3pt-rec}

In momentum space, the special conformal WT identity is stated as
\begin{align}
\label{SCWT}
\bsb.\bsK \, \lan O_1(\bsk_1)O_2(\bsk_2)\ep^s.O_3(\bsk_3) \ran' = 0\,,
\end{align}
where an explicit form of the special conformal generator $\bsK$ is given by
\begin{align}
\bsb.{\bsK}&=\sum_{i=1}^3
\Big[
\bsb.\bspd_i\big(-2(\De_i-d+1)+2\bsk_i.\bspd_i\big)-(\bsb.\bsk_i)\bspd_i^2
\Big] \nn\\
\label{SC_generator}
& \qquad
+2\Big[
\bsep.\bspd_3(\bsb.\bspd_\bsep)
-(\bsb.\bsep)(\bspd_3.\bspd_\bsep)
\Big]\,.
\end{align}
For later convenience, we contracted the generator with an arbitrary vector $\bsb$. Also, we introduced a shorthand notation,
\begin{align}
\bspd_i=\frac{\partial}{\partial \bsk_i}\,,
\quad
\bspd_\bsep=\frac{\partial}{\partial \bsep}\,.
\end{align}
A useful observation is that the second line of Eq.~\eqref{SC_generator} trivially acts on $\bsk_3$-independent functions. We therefore write the
three-point functions as functions of $\bsk_1$ and $\bsk_2$. In particular, $k_3$ should be understood as $k_3^2=(\bsk_1+\bsk_2)^2$. Another observation is that the last term in the first line of Eq.~\eqref{SC_generator} vanishes when $\bsb.\bsk_1=\bsb.\bsk_2=0$. For this choice of $\bsb$ (we also assume $\bsep.\bsb\neq0$ for later convenience), the special conformal WT identity is simply~\cite{Arkani-Hamed:2015bza}
\begin{align}
\label{SCWT_3pt}
\sum_{i=1}^2
\Big[
\bsb.\bspd_i\big(-2(\De_i-d+1)
+2\bsk_i.\bspd_i\big)
\Big]
\lan O_1(\bsk_1)O_2(\bsk_2)\ep^s.O_3(\bsk_3) \ran' = 0\,.
\end{align}
Notice that this is a subset of the special conformal WT identity, hence there still exist two independent conditions associated with $\bsb$ which does not satisfy $\bsb.\bsk_1=\bsb.\bsk_2=0$. These two conditions may be used to constrain $a_0$ as we discuss in Appendix~\ref{App-subsec:WTa0}.

\medskip
We then apply the condition~\eqref{SCWT_3pt} to the ansatz~\eqref{00s_ansatz}. By reorganizing the differential operator in terms of $\bsb.\bspd_1\pm\bsb.\bspd_2$, Eq.~\eqref{SCWT_3pt} can be rephrased as
\begin{align}
&{}
\Big[
(\De_3-2)(\bsb.\bspd_1+\bsb.\bspd_2)
+ (\bsb.\bspd_1-\bsb.\bspd_2)
\big(-(\De_1-\De_2)+\bsk_1.\bspd_1-\bsk_2.\bspd_2\big)
\Big] \nn\\
&{} \qquad
\times \, \Big[ k_3^{\Delta_t-2d-s}\sum_{n=0}^s\gamma^{s-n}\delta^n a_n(\kappa_1,\kappa_2) \Big] = 0
\,,
\label{SCWT-k}
\end{align}
where we used the fact that three-point functions have a scaling dimension $\Delta_t-2d$ (after dropping the delta function for momentum conservation).
The next step is to rewrite the momentum derivatives as those in $\ga,\de,\ka_1,\ka_2$.
For this, we use the following relations:
\begin{align}
&
\bsb.\bspd_i f(k_1,k_2,k_3)=0
\quad
(i=1,2)\,,
\nn\\
&
(\bsb.\bspd_1-\bsb.\bspd_2)\gamma=2(\bsep.\bsb)\,,
\quad
(\bsb.\bspd_1-\bsb.\bspd_2)\delta=0\,,
\nn\\
&
(\bsb.\bspd_1+\bsb.\bspd_2)\gamma=0\,,
\quad
(\bsb.\bspd_1+\bsb.\bspd_2)\delta=2(\bsep.\bsb)\,,
\nn\\
&
(\bsk_1.\bspd_1-\bsk_2.\bspd_2)\gamma=\delta\,,
\quad
(\bsk_1.\bspd_1-\bsk_2.\bspd_2)\delta=\gamma\,,
\nn\\
&(\bsk_1.\bspd_1-\bsk_2.\bspd_2)k_3=(\kappa_1^2-\kappa_2^2)k_3\,,
\nn\\
&
(\bsk_1.\bspd_1-\bsk_2.\bspd_2)\kappa_1=-(\kappa_1^2-\kappa_2^2-1)\kappa_1\,,
\nn\\
&
(\bsk_1.\bspd_1-\bsk_2.\bspd_2)\kappa_2=-(\kappa_1^2-\kappa_2^2+1)\kappa_2
\,.
\end{align}
An important point is that the operator $\bsb.\bspd_i$ trivially acts on $a_n$ since this is a function of $k_1$, $k_2$, and $k_3$. We then arrive at
\begin{align}
\label{SCWT-gammadeltakappa}
0=
\Big[
(\De_3-1)\partial_\delta
+\delta\partial_\gamma^2
+\gamma\partial_\gamma\partial_\delta
-\partial_\gamma{\cD}
\Big]\sum_{n=0}^s\gamma^{s-n}\delta^na_n(\kappa_1,\kappa_2)\,.
\end{align}
Here we used $\bsep.\bsb\neq0$ and introduced a differential operator,
\begin{align}
\label{cD_generalD}
{\cD}
&=(\kappa_1^2-\kappa_2^2)\left(
\kappa_1\partial_{\kappa_1}+\kappa_2\partial_{\kappa_2}
-\De_t+s+2d
\right)
-\Big(\kappa_1\partial_{\kappa_1}-\kappa_2\partial_{\kappa_2}-\De_1+\De_2\Big)\,,
\end{align}
which contains derivatives in $\ka_1$ and $\ka_2$ only. Also note that we regard $\gamma$, $\delta$, $\kappa_1$ and $\kappa_2$ as independent variables in the expressions~\eqref{SCWT-gammadeltakappa}-\eqref{cD_generalD}. For example, we have $\partial_{\kappa_i}\gamma=0$ and $[\cD,\gamma]=0$, which let us solve the recursion relation as we discuss in the next subsection.

\medskip
Finally, by rewriting Eq.~\eqref{SCWT-gammadeltakappa} into the form,
\begin{align}
&{}
\sum_{n=0}^s
\Big[
-(s-n)\gamma^{s-n-1}\delta^{n}\,{\cD} \nn\\
&{} \qquad\quad 
+n(\De_3-1+s-n)\gamma^{s-n}\delta^{n-1}
+(s-n)(s-n-1)\gamma^{s-n-2}\delta^{n+1}
\Big]
a_n=0\,,
\end{align}
we may pick up the desired recursion relation for $a_n$ from the $\mathcal{O}(\delta^{n})$ terms as
\begin{align}
a_{n+1}=\frac{s-n}{(n+1)(\De_3-2+s-n)}
\big({\cD}a_n-(s-n+1)a_{n-1}\big)\,,
\end{align}
where we define $a_{-1}=0$ for convenience. Also note that this recursion relation is consistent with $a_{s+1}=0$. Once a concrete form of $a_0$ is given, we may compute $a_n$ at least recursively.
We can easily see that $a_n$ is schematically of the form,
\begin{align}
\label{an-in-D1}
a_{n}\sim
\sum_{k=0}^{[n/2]} \, {\cD}^{n-2k}a_0\,,
\end{align}
up to numerical coefficients independent of momenta and polarization vectors. 
Substituting this back into \eqref{00s_ansatz}, we have
\begin{align}
\label{3pt-spin-with-gamma-delta-D-nk}
\lan O_1(\bsk_1)O_2(\bsk_2)\ep^s.O_3(\bsk_3) \ran'
&=k_3^{\Delta_t-2d-s}\delta^s\sum_{n=0}^s\sum_{k=0}^{[n/2]}
\al_{n,k}(\ga/\de)^{s-n}{\cD}^{n-2k}a_0\,,
\end{align}
where we wrote numerical coefficients as $\al_{n,k}$. Note that $\al_{0,0}=1$ in particular.

\subsection{A closed form of 3pt functions}
\label{App-subsec:closed-form}
It might seem difficult to find a closed form of $\al_{n,k}$, but it is actually possible to find a closed form of \eqref{3pt-spin-with-gamma-delta-D-nk} by solving the special conformal WT identity~\eqref{SCWT-gammadeltakappa}. 
The special conformal WT identity in terms of \eqref{3pt-spin-with-gamma-delta-D-nk} reads
\begin{align}
\Big[
(\De_3-1)\partial_\delta
+\delta\partial_\gamma^2
+\gamma\partial_\gamma\partial_\delta \,
-\partial_\gamma{\cD}
\Big]\Big[\delta^s\sum_{n=0}^s\sum_{k=0}^{[n/2]}
\al_{n,k}(\ga/\de)^{s-n}{\cD}^{n-2k}a_0\Big]=0
\,.
\label{SCWT-gammadeltakappa2}
\end{align}
It is convenient to further rewrite it as
\begin{align}
\Big[
(\De_3-1)\partial_\delta
+\delta\partial_\gamma^2
+\gamma\partial_\gamma\partial_\delta \,
-\partial_\gamma{\cD}
\Big]\Big[\delta^sh_{\cD}(\gamma/\delta)a_0\Big]=0\,,
\label{SCWT-gammadeltakappa3}
\end{align}
where we introduced
\begin{align}
\label{h-def}
h_{\cD}(z)=\sum_{n=0}^s\sum_{k=0}^{[n/2]}\al_{n,k}z^{s-n}{\cD}^{n-2k}
\,.
\end{align}
Let us here recall that we regard $\gamma$, $\delta$ and $\kappa_i$ as independent variables, so that $[\cD,\gamma]=[\cD,\delta]=0$, in the relation~\eqref{SCWT-gammadeltakappa3}. Thanks to this property, we can think of $h_{\cD}(z)$ as a function of $z$ with a parameter $\cD$. Our problem is now equivalent to solving the equation,
\begin{align}
\label{diffeq-hD}
0=
\Big[
(\De_3-1)\partial_\de
+\de\partial_\ga^2
+\ga\partial_\ga\partial_\de
-\partial_\ga{\cD}
\Big]\Big[\de^s\,{h}_{\cD}(\ga/\de)\Big]\,,
\end{align}
for $\gamma$ and $\delta$.
By setting $z=\ga/\de$ in the above, we can reduce it into
\begin{align}
\Big[\,s(\De_3-1)
 + \big((s-\De_3)z-{\cD}\big)\partial_z  + (1-z^2)\partial_z^2\, \Big]h_{\cD}(z)=0\,.
\end{align}
Notice that the function $h_{\cD}(z)$, by construction \eqref{h-def}, is a polynomial in $z$ of order $s$. We can therefore find its unique solution, which is a hypergeometric function,
\begin{align}
h_{\cD}(z)&\propto 
\,_2F_1\bigg( -s, \De_3-1;\frac{\De_3-s+\cD}{2};\frac{1-z}{2} \bigg)\,.
\end{align}
This is indeed a polynomial of order $s$ because the hypergeometric function has negative integer $-s$ in its first slot.\footnote{
The concrete relation is 
\begin{align}
\label{2F1-s}
\,_2F_1(-s, b; c; z) = \sum_{n=0}^s (-)^n \binom{s}{n} 
\frac{(b)_n}{(c)_n}z^n.
\end{align}}
The coefficient of proportionality may depend on $\cD$, but we can fix it by requiring $h_{\cD}(z)=z^s+\cdots$, which follows from $\al_{0,0}=1$ mentioned in the previous subsection. We then arrive at
\begin{align}
h_{\cD}(z)
&=
\sum_{n=0}^s
\frac{2^{s-n}(\frac{\De_3-s+\cD}{2}+n)_{s-n}}{(\De_3-1+n)_{s-n}}
\binom{s}{n}
\left(z-1\right)^n\,.
\end{align}
By substituting this back into \eqref{3pt-spin-with-gamma-delta-D-nk}, we obtain
\begin{align}
&\lan O_1(\bsk_1)O_2(\bsk_2)\ep^s.O_3(\bsk_3) \ran'
\nn\\*
&\quad =k_3^{\Delta_t-2d-s}\sum_{n=0}^s
2^{s-n}\binom{s}{n}
\left(\gamma-\delta\right)^n\delta^{s-n}
\frac{\left(\frac{\De_3-s+{\cD}}{2}+n\right)_{s-n}}{(\De_3-1+n)_{s-n}}
\,a_0
\nn\\*
&\quad =k_3^{\Delta_t-2d-s}\sum_{n=0}^s
\frac{s!}{n!(s-n)!}
\left(\bsep.\bsk_2\right)^{s-n}(\bsep.\bsk_3)^{n}
\frac{\left(\frac{\De_3+s+{\cD}}{2}-n\right)_{n}}{(\De_3-1+s-n)_{n}}
\,(-2)^{s}a_0
\,,
\label{3pt-spin-with-gamma-delta-D-nk-a0}
\end{align}
where we performed a change $n\to s-n$ at the second equality.

\medskip
Our final task is now to find the initial condition $a_0$ for the recursion relation. We work on this problem in the next subsection and find
\begin{align}
\label{a_0}
{(-2)^{s}}a_0={C_{123}}\int_0^\infty \frac{dz}{z^{d+1}}z^s\mathcal{B}_{\nu_1}(\kappa_1;z)\mathcal{B}_{\nu_2}(\kappa_2;z)\mathcal{B}_{\nu_3}(1;z)\,,
\end{align}
where the OPE coefficient $C_{123}$ in momentum space is related to the position space one as~Eq.~\eqref{OPE_sst_mp}. Also its normalization is chosen in such a way that the final expression is simple. 
By substituting this explicit form into Eq.~\eqref{3pt-spin-with-gamma-delta-D-nk-a0} and rescaling the integration variable in the triple-$K$ integral as $z\to k_3z$, we conclude that
\begin{align}
&\lan O_1(\bsk_1)O_2(\bsk_2)\ep^s.O_3(\bsk_3) \ran'
=C_{123}\sum_{n=0}^s
\frac{s!}{n!(s-n)!}
\left(\bsep.\bsk_2\right)^{s-n}(\bsep.\bsk_3)^{n}\nn\\
\label{00s_momentum_nice}
&\qquad\qquad\qquad\qquad \times
\frac{\big(\frac{\De_3+s+{\fkD_{123}}}{2}-n\big)_{n}}{(\De_3-1+s-n)_{n}}
\int_0^\infty \frac{dz}{z^{d+1}}z^s\mathcal{B}_{\nu_1}(k_1;z)\mathcal{B}_{\nu_2}(k_2;z)\mathcal{B}_{\nu_3}(k_3;z)
\,,
\end{align}
where we introduced the rescaled differential operator ${\fkD_{123}}$ as\footnote{We added the subscripts $123$ (the labels of the operators $O_i$) to the differential operator $\fkD$ in order to clarify the dependence on the momenta $k_i$ and the scaling dimensions $\De_i$ ($i=1,2,3$), and the spin $s$ of $O_3$.}
\begin{align}
\label{fkD_generalD}
\fkD_{123}
&=\frac{k_1^2-k_2^2}{k_3^2}\left(
k_1\partial_{k_1}+k_2\partial_{k_2}
-\De_t+s+2d
\right)
-\Big( k_1\partial_{k_1}-k_2\partial_{k_2}-\De_1+\De_2 \Big) \,.
\end{align}
This is the first closed form~\eqref{3pt_spin_generealD}. Here 
$k_1$, $k_2$ and $k_3$ should be understood as independent variables when ${\fkD_{123}}$ acts on the triple-$K$ integral. More concretely, $\pd_{k_i}k_3=0$ $(i=1,2)$ for example.
Note also that the final form \eqref{00s_momentum_nice} depends on the space dimension $d$ only through the differential operator $\fkD_{123}$ and the triple-K integral.

\medskip
Before discussing the initial condition problem, let us take a quick look at the symmetry under the exchange $1 \leftrightarrow 2$ of the two scalars.
Since Eq.~\eqref{3pt-spin-with-gamma-delta-D-nk} implies the invariance of three-point functions under $(\delta,{\cD})\to(-\delta,-{\cD})$, we may reformulate Eq.~\eqref{3pt-spin-with-gamma-delta-D-nk-a0} as
\begin{align}
&\lan O_1(\bsk_1)O_2(\bsk_2)\ep^s.O_3(\bsk_3) \ran'
\nn\\*
&=k_3^{\Delta_t-6-s}\sum_{n=0}^s
2^{s-n}\binom{s}{n}
\left(\gamma+\delta\right)^n(-\delta)^{s-n}
\frac{\left(\frac{\De_3-s-{\cD}}{2}+n\right)_{s-n}}{(\De_3-1+n)_{s-n}}
\,a_0
\,.
\end{align}
This yields another expression for three-point functions:
\begin{align}
&\lan O_1(\bsk_1)O_2(\bsk_2)\ep^s.O_3(\bsk_3) \ran'
=(-)^sC_{123}\sum_{n=0}^s
\frac{s!}{n!(s-n)!}
\left(\bsep.\bsk_1\right)^{s-n}(\bsep.\bsk_3)^{n}\nn\\
\label{00s_momentum_nice_another}
&\qquad\qquad\qquad\qquad \times
\frac{\big(\frac{\De_3+s-{\fkD_{123}}}{2}-n\big)_{n}}{(\De_3-1+s-n)_{n}}
\int_0^\infty \frac{dz}{z^{d+1}}z^s\mathcal{B}_{\nu_1}(k_1;z)\mathcal{B}_{\nu_2}(k_2;z)\mathcal{B}_{\nu_3}(k_3;z)
\,.
\end{align}
By comparing this expression with Eq.~\eqref{00s_momentum_nice}, we see in particular that the three-point function vanishes when two scalars are identical and the tensor has an odd spin $s$.

\subsection{Initial condition $a_0$}
\label{App-subsec:WTa0}

So far, we have used a subset of the special conformal WT identity~\eqref{SCWT} with $\bsb$ satisfying $\bsb.\bsk_1=\bsb.\bsk_2=0$.
In this subsection, we determine the initial condition $a_0$ by solving the residual WT identities for $\bsb.\bsk_1\neq0$ and $\bsb.\bsk_2\neq0$. To simplify the calculation, we also set $\bsb.\bsep=0$~\cite{Arkani-Hamed:2015bza}.
For later convenience, let us first modify the ansatz~\eqref{00s_ansatz} as
\begin{align}
\lan O_1(\bsk_1)O_2(\bsk_2)\ep^s.O_3(\bsk_3) \ran'
&=\sum_{n=0}^s\gamma^{s-n}\delta^n \widetilde{a}_n(k_1,k_2,k_3)
\,,
\end{align}
where we rescaled $a_n$ as $\wt{a}{}_n(k_1,k_2,k_3)=k_3^{\De_t-2d-s}a_n({k_1}/{k_3}, {k_2}/{k_3})$.
Also, it follows from the dilatation WT identity that $\widetilde{a}_n$'s are $(\De_t-2d-s)$-th order homogeneous polynomials.
The special conformal WT identity for $\bsb.\bsk_1\neq0$, $\bsb.\bsk_2\neq0$ and $\bsb.\bsep=0$ is then given by 
\begin{align}
\label{bK-k1k2k3}
\bsb.\bsK  \sum_{n=0}^s \ga^{s-n}\de^n \, \wt{a}{}_n(k_1,k_2,k_3) 
= 0\,,
\end{align}
where the generator $\bsK$ in \eqref{bK-k1k2k3} reads\footnote{
When deriving Eq.~\eqref{SCK}, it is convenient to identify $k_3$ as $k_3=|\bsk_3|$, rather than $k_3=|\bsk_1+\bsk_2|$.}
\begin{align}
\bsb.\bsK &= \bsb.\bsk_1 
\Big( -\cK_1(\nu_1)+\cK_3(\nu_3)
+\frac{2}{k_3}\frac{\pd}{\pd k_3}\de\pd_\de
+\frac{2}{k_3}\frac{\pd}{\pd k_3}\de\pd_\ga \Big)
\nn\\
\label{SCK}
&{} \quad
+ \bsb.\bsk_2
\Big( -\cK_2(\nu_2)+\cK_3(\nu_3)+\frac{2}{k_3}\frac{\pd}{\pd k_3}\de\pd_\de-\frac{2}{k_3}\frac{\pd}{\pd k_3}\de\pd_\ga \Big)\,,
\end{align}
with a differential operator,
\begin{align}
\cK_i(\nu_i) = \frac{\pd^2}{\pd k_i^2} - \frac{2\nu_i-1}{k_i}\frac{\pd}{\pd k_i}.
\end{align}
Also note that, similar to \eqref{SCWT-gammadeltakappa}, we regard $\ga,\de,k_1,k_2,k_3$ as independent variables.
Picking up $\de^0$-terms gives closed equations for $\widetilde{a}_0$,
\begin{align}
\label{SCWTa0}
\cK_1(\nu_1)\wt{a}{}_0=\cK_2(\nu_2)\wt{a}{}_0=\cK_3(\nu_3)\wt{a}{}_0.
\end{align}
This is actually the same as the special conformal WT identity for scalar three-point functions. Following the scalar three-point function case~\cite{Bzowski:2013sza}, let us take the ansatz,
\begin{align}
\label{a0_ansatz}
\wt{a}{}_0\propto\int_0^\infty\frac{dz}{z} z^{s+2d-\Delta_t}
f_1(k_1z)f_2(k_2z)f_3(k_3z)\,,
\end{align}
where the proportionality constant is $k_i$-independent. The integral~\eqref{a0_ansatz} gives a $(\Delta_t-2d-s)$-th order homogeneous polynomial in $k_i$, so that it satisfies the dilatation WT identity manifestly. The special conformal WT identity~\eqref{SCWTa0} may then be rephrased as
\begin{align}
\cK_i(\nu_i)f_i(k_iz)=z^2f_i(k_iz)\,.
\end{align}
Here we have fixed the normalization of the r.h.s. by rescaling of $z$, which may be absorbed into the proportionality constant in Eq.~\eqref{a0_ansatz}. Its general solution is given by
\begin{align}
f_i(x)
=x^{\nu_i}\Big[
A\,K_{\nu_i}(x)
+B\,I_{\nu_i}(x)
\Big]\,.
\end{align}
Similarly to the argument at the end of Sec.~\ref{subsec:factorization3pt},
the $B$ component leads to undesired singularities of the integral~\eqref{a0_ansatz}, hence we set $B=0$. This choice means that $f_i(k_iz)$ is a bulk-to-boundary propagator multiplied by a factor $z^{\nu}$, up to a normalization constant. Therefore, $\widetilde{a}_0$ can be written as
\begin{align}
\widetilde{a}_0(k_1,k_2,k_3)\propto\int_0^\infty \frac{dz}{z^{d+1}}z^s\mathcal{B}_{\nu_1}(k_1;z)\mathcal{B}_{\nu_2}(k_2;z)\mathcal{B}_{\nu_3}(k_3;z)\,,
\end{align}
and $a_0(\kappa_1,\kappa_2)=\widetilde{a}_0(\kappa_1,\kappa_2,1)$ by definition. Notice here that Eq.~\eqref{a_0} corresponds to the choice of the proportionality constant $(-2)^{-s}C_{123}$.

\subsection{Three-point functions by Fourier transformation}
\label{App-subsec:3pt-FT}

Finally, let us discuss the relation between the momentum space correlator and the position space one.
The Fourier transformation of the position space correlator~\eqref{3pt-s-position} is given by
\begin{align}
&{}
\lan O_1(\bsk_1)O_2(\bsk_2){\ep}^s.O_3(\bsk_3) \ran'
\nn\\
\label{Fourier-1}
&{} \quad
= \widetilde{C}_{123} \sum_{m_1+m_2=s}(-)^{m_2}i^s
\binom{s}{m_1}
\left( \bsep.\frac{\pd}{\pd \bsk_1} \right)^{m_1} 
\left( \bsep.\frac{\pd}{\pd \bsk_2} \right)^{m_2}
\nn\\
&{} \quad\quad
\times \int d^dy_{1} d^dy_{2} \, 
\frac{e^{-i\bsk_1.\bsy_{1}-i\bsk_2.\bsy_{2}}}
{[(\bsy_{1}-\bsy_{2})^2]^{\frac{1}{2}(\De_{123}+s)}
[y_1^2]^{\frac{1}{2}(\De_{231}+s-2m_1)}
[y_2^2]^{\frac{1}{2}(\De_{312}+s-2m_2)}}\,,
\end{align}
where we replaced the position coordinates in the numerator of Eq.~\eqref{3pt-s-position} by momentum derivatives, after the binomial expansion.
The integral here is essentially the same as the scalar three-point function, hence it gives a triple-$K$ integral. A more precise relation is given through a formula (see, e.g.,~\cite{Bzowski:2013sza} for derivation),
\begin{align}
\label{tripleK-Fourier}
&{}
\int\!d^dy_1 d^dy_2 \frac{e^{-i\bsk_1.\bsy_1-i\bsk_2.\bsy_2}}
{[(\bsy_1-\bsy_2)^2]^{a_3} [y_2^2]^{a_1} [y_1^2]^{a_2}}
\nn\\
&
= \frac{2^{4+3h-2a_t} \pi^{2h}}{\Ga(a_1)\Ga(a_2)\Ga(a_3)\Ga(a_t-h)}
\int_0^\infty\! \frac{dz}{z^{1-h}} \prod_{i=1}^3 \Big\{ k_i^{a_t-a_i-h}K_{a_t-a_i-h}(k_iz) \Big\}\Big|_{k_3=|\bsk_1+\bsk_2|}\,,
\end{align}
where we defined $a_t=a_1+a_2+a_3$.
Substituting (B.42) back into (B.41), we find the following derivatives:
\begin{align}
\label{3pt-spin-FT-mid}
\left( \bsep.\frac{\pd}{\pd \bsk_1} \right)^{m_1} 
\left( \bsep.\frac{\pd}{\pd \bsk_2} \right)^{m_2}
\int_0^\infty \!\frac{dz}{z^{1-h}} \prod_{i=1}^3 \Big\{ k_i^{\nu_i+m_i}K_{\nu_i+m_i}(k_iz) \Big\}\Big|_{k_3=|\bsk_1+\bsk_2|}\,,
\end{align}
where we introduced $m_3=0$ for notational simplicity.
Since the triple-$K$ integral depends only on the magnitudes $k_i$ of the momenta $\bsk_i$, we can replace the derivatives in $\bsk_i$ by those in $k_i$. More explicitly, for $i=1,2$ we have
\begin{align}
\left( \bsep.\frac{\pd}{\pd\bsk_i} \right)^a f(k_i) 
&= (\bsep.\bsk_i)^a \left( \frac{1}{k_i}\frac{d}{d k_i} \right)^a f(k_i), \\
\left( \bsep.\frac{\pd}{\pd\bsk_i} \right)^a f(|\bsk_1+\bsk_2|) 
&= (-\bsep.\bsk_3)^a \left( \frac{1}{k_3}\frac{d}{d k_3} \right)^a f(k_3) 
\bigg|_{\bsk_3=-\bsk_1-\bsk_2}, 
\end{align}
where each function $f$ depends only on the magnitudes of the momenta. Also, we used $\bsep.\bsep=0$.
Combining them with the generalized Leibniz rule ($i=1,2$), we have 
\begin{align}
&{}
\left( \bsep.\frac{\pd}{\pd \bsk_i} \right)^{m_i} \big[ f_i(k_i) f_3(|\bsk_1+\bsk_2|) \big]
\nn\\
&{} \quad
= \sum_{a=0}^{m_i}
\binom{m_i}{a}
\Bigg[ \left( \bsep.\frac{\pd}{\pd \bsk_i} \right)^{m_i-a} f_i(k_i) \Bigg] \times
\Bigg[ \left( \bsep.\frac{\pd}{\pd \bsk_i} \right)^{a} f_3(|\bsk_1+\bsk_2|) \Bigg]\,.
\end{align}
Eq.~\eqref{3pt-spin-FT-mid} may then be reformulated as
\begin{align}
&\sum_{m_1+m_2=s} \sum_{q_1=0}^{m_1} \sum_{q_2=0}^{m_2}
\binom{m_1}{q_1}
\binom{m_2}{q_2}
(\bsep.\bsk_1)^{m_1-q_1} (\bsep.\bsk_2)^{m_2-q_2} (-\bsep.\bsk_3)^{q_1+q_2}
\nn\\
&\quad
\times\int_0^\infty \!\frac{dz}{z^{1-h}} (-z)^s\prod_{i=1}^2 \Big\{ k_i^{\nu_i+q_i}K_{\nu_i+q_i}(k_iz) \Big\}\,
k_3^{\nu_3-q_1-q_2}K_{\nu_3-q_1-q_2}(k_3z)
\,.
\end{align}
Here and in the rest of this section, we use $k_3$ and $|\bsk_1+\bsk_2|$ interchangeably because there are no more derivatives.
Also we used a relation following from Eq.~\eqref{Bessel_useful},
\begin{align}
\frac{1}{k}\frac{\pd}{\pd k}\big( k^\nu K_\nu(kz) \big) = -zk^{\nu-1}K_{\nu-1}(kz)\,,
\end{align}
which induced the shift of the arguments in the triple-$K$ integrals.
By applying this result to \eqref{3pt-spin-FT-mid}, we arrive at the second expression for momentum space correlators,
\begin{align}
\label{3pt-spin-mom1}
\nonumber
&\lan O_1(\bsk_1)O_2(\bsk_2)\ep^s.O_3(\bsk_3) \ran'
\\
& \quad
= c_{123} 
\sum_{m_1+m_2=s} \sum_{q_1=0}^{m_1} \sum_{q_2=0}^{m_2}
d_{123}(q_1,q_2,m_1,m_2)\,(-)^{m_2}
\nn\\
&\qquad\quad\,\,\times
\, (\bsep.\bsk_1)^{m_1-q_1} (\bsep.\bsk_2)^{m_2-q_2} \,(-\bsep.\bsk_3)^{q_1+q_2}
\cI_{s}(\nu_1+q_1, \, \nu_2+q_2, \, \nu_3-q_1-q_2)\,,
\end{align}
where we wrote the triple-$K$ integral in terms of the bulk-to-boundary propagator $\mathcal{B}_{\nu}$ as
\begin{align}
&\cI_s(\al_1,\al_2,\al_3)
= \int_0^\infty \frac{dz}{z^{d+1}}z^s\mathcal{B}_{\al_1}(k_1;z)\mathcal{B}_{\al_2}(k_2;z)\mathcal{B}_{\al_3}(k_3;z)\,.
\end{align}
We also introduced a new overall factor,
\begin{align}
\label{c123}
&c_{123}=\wt{C}{}_{123} \, 2^{1-s} \pi^d i^s (-)^s 
\frac{\Gamma(\nu_1)\Gamma(\nu_2)\Gamma(\nu_3)}
{\Ga(\frac{\De_{123}+s}{2}) \Ga(\frac{\De_{231}+s}{2}) \Ga(\frac{\De_{312}+s}{2}) \Ga(\frac{\De_t+s-d}{2})}\,,
\end{align}
and a numerical coefficient,
\begin{align}
\label{A123}
&d_{123}(q_1,q_2,m_1,m_2) =
\frac{s!}{q_1!q_2!(m_1-q_1)!(m_2-q_2)!} \frac{(\nu_1)_{q_1}(\nu_2)_{q_2}}{(\nu_3-q_1-q_2)_{q_1+q_2}}
\nn\\
&{} \qquad\qquad\qquad\qquad\quad\quad \times
\Big( \frac{\De_{231}+s}{2}-m_1 \Big)_{m_1} \Big( \frac{\De_{132}+s}{2}-m_2 \Big)_{m_2}\,.
\end{align}

\paragraph{Initial value $a_0$}

Let us now clarify the relation between the normalization factor $C_{123}$ in momentum space and the position space OPE coefficient $\widetilde{C}_{123}$. For this purpose, let us rewrite the closed form \eqref{3pt-spin-mom1} into the form of the ansatz~\eqref{00s_ansatz} as
\begin{align}
&\lan O_1(\bsk_1)O_2(\bsk_2)\ep^s.O_3(\bsk_3) \ran'
\nn\\
&
= c_{123} k_3^{\De_t-2d-s} 
\sum_{m_1+m_2=s} \sum_{q_1=0}^{m_1} \sum_{q_2=0}^{m_2}d_{123}(q_1,q_2,m_1,m_2) \, (-)^{m_2}\,2^{-s+q_1+q_2} \nn\\
&\,\,~\times
(\gamma+\delta)^{m_1-q_1} (\delta-\gamma)^{m_2-q_2} \,\delta^{q_1+q_2}
\,
\cI_{s}(\nu_1+q_1, \, \nu_2+q_2, \, \nu_3-q_1-q_2)\Big|_{k_i=\kappa_i,\,k_3=1}\,.
\end{align}
Comparing this with Eq.~\eqref{00s_ansatz}, we find
\begin{align}
\label{a0}
a_0
&=2^{-s}c_{123}\sum_{m_1+m_2=s}d_{123}(0,0,m_1,m_2)\,\cI_{s}(\nu_1, \, \nu_2, \, \nu_3)\Big|_{k_i=\kappa_i,\,k_3=1}\,.
\end{align}
The normalization factor $C_{123}$ in Eq.~\eqref{a_0} is therefore given by
\begin{align}
\label{C_pre}
C_{123}=(-1)^s\,c_{123}\sum_{m_1+m_2=s}d_{123}(0,0,m_1,m_2)\,.
\end{align}
More explicitly, we may use the formulae \eqref{2F1-s} and
\begin{align}
{}_2F_1(\al,\bt;\ga;1) = \frac{\Ga(\ga)\Ga(\ga-\al-\bt)}{\Ga(\ga-\al)\Ga(\ga-\bt)}\,,
\end{align}
to evaluate the summation in Eq.~\eqref{C_pre} as\footnote{
One might wonder that the exchange symmetry of two scalars is not manifest in the second expression. However, we may also write
\begin{align}
\sum_{m_1+m_2=s}d_{123}(0,0,m_1,m_2)
&
=(\tfrac{\De_{132}-s}{2})_s \, {}_2F_1\big(-s,1-\tfrac{\De_{231}+s}{2};\tfrac{\De_{132}-s}{2};1\big)\,,
\end{align}
instead of the first equality in Eq.~\eqref{d_sum}. Both the two expressions indeed give the same result at the end.}
\begin{align}
\sum_{m_1+m_2=s}d_{123}(0,0,m_1,m_2)
&
=(\tfrac{\De_{231}-s}{2})_s \, {}_2F_1\big(-s,1-\tfrac{\De_{132}+s}{2};\tfrac{\De_{231}-s}{2};1\big)
\nn\\
\label{d_sum}
&=(\De_3-1)_s
\,.
\end{align}
We therefore conclude that
\begin{align}
\nonumber
C_{123}&=(-1)^s(\De_3-1)_s\,c_{123}
\\
\label{OPE_sst_mp}
&=\wt{C}{}_{123} \, 2^{1-s} \pi^d i^s (\De_3-1)_s
\frac{\Gamma(\nu_1)\Gamma(\nu_2)\Gamma(\nu_3)}
{\Ga(\frac{\De_{123}+s}{2}) \Ga(\frac{\De_{231}+s}{2}) \Ga(\frac{\De_{312}+s}{2}) \Ga(\frac{\De_t+s-d}{2})}\,.
\end{align}

\section{Explicit form of $A_{123_m}$}
\label{app:A123m}

In this appendix we derive the analytic form \eqref{00s_helicity}-\eqref{A_123m} of the three-point function with helicity $m$ tensor $O_{3_m}$. The central part of the computation is the integral,
\begin{align}
\label{fourier-epk2-epk3}
\int_0^{2\pi} \frac{d\psi}{2\pi} \, e^{-im\psi} (\bsep.\bsk_2)^{s-n} (\bsep.\bsk_3)^n \,.
\end{align}
Under the parametrizations \eqref{k3ep}-\eqref{k2}, it is given by
\begin{align}
\eqref{fourier-epk2-epk3}&=
(k_2)^{s-n}(ik_3)^n\int_0^{2\pi} \frac{d\psi}{2\pi} \, e^{-im\psi}\, [\cos(\psi-\chi)\sin\te+i\cos\te]^{s-n}
\nn\\*
&=(k_2)^{s-n}(ik_3)^ne^{-im\chi}\int_0^{2\pi} \frac{d\psi}{2\pi} \, e^{-im\psi}\, [\cos\psi\sin\te+i\cos\te]^{s-n}
\nn\\*
\label{c2}
&=(k_2)^{s-n}(ik_3)^ne^{-im\chi}\int_0^{2\pi} \frac{d\psi}{2\pi} \, \cos {|m|}\psi\cdot [\cos\psi\sin\te+i\cos\te]^{s-n}
\,,
\end{align}
where we performed a change of the integration variable $\psi\to\psi+\chi$ at the second equality.

\medskip
To perform the integral in the last line of Eq.~\eqref{c2},
let us first set $t=\cos\psi$ and use the Chebyshev polynomial \eqref{Chebyshev}. Applying the Rodrigues formula \eqref{rodrigues} to the Chebyshev polynomial, we can rewrite the integral as
\begin{align}
\label{lmm-t}
\frac{(-1)^{|m|}}{2^{|m|}(1/2)_{|m|}\pi}
\int_{-1}^1 dt \, \bigg[ \frac{d^{|m|}}{dt^{|m|}}(1-t^2)^{{|m|}-\frac{1}{2}} \bigg] \cdot 
( \, t\sin\te+i\cos\te \, )^{s-n} \,.
\end{align}
Integrating this by parts ${|m|}$ times, we find 
\begin{align}
\label{lmm-Legendre}
\eqref{lmm-t}
&= \frac{\sin^{|m|}\te}{2^{|m|}(1/2)_{|m|}\pi} \frac{(s-n)!}{(s-n-{|m|})!}
\int_{-1}^1 dt \, (1-t^2)^{{|m|}-\frac{1}{2}} \cdot 
( \, t\sin\te+i\cos\te \, )^{s-n-|m|} \,.
\end{align}
Note that it vanishes when $s-n-|m|<0$. To evaluate the integral in the above, we use the following integral form of the Gegenbauer polynomial \cite{AH}:\footnote{
A derivation of the integral formula will be provided also in our forthcoming paper~\cite{crossing2}.}
\begin{align}
\label{G_formula}
\frac{C^{(D/2-1)}_{n}(\cos\te)}{C^{(D/2-1)}_{n}(1)}
= \frac{{\rm vol}(S^{D-3})}{{\rm vol}(S^{D-2})}
\int_{-1}^1 dt \, (1-t^2)^{\frac{D-4}{2}} (\cos\te - it\sin\te)^n \,,
\end{align}
where $C^{(\al)}_n$ is the Gegenbauer polynomial \eqref{Gegenbauer} and ${\rm vol}(S^D)$ is the volume of the $D$-dimensional unit sphere 
\begin{align}
{\rm vol}(S^D) = \frac{2\pi^{\frac{D+1}{2}}}{\Ga(\frac{D+1}{2})} \,.
\end{align}
It then follows from Eqs.~\eqref{lmm-Legendre} and~\eqref{G_formula} that
\begin{align}
\eqref{lmm-t}
&= \frac{i^{s-n-{|m|}} (s-n)!}{2^{|m|} (s-n-{|m|})! {|m|}!}
 \widehat{P}_{s-n,{|m|}}(\cos\te) \,.
\end{align}
Here we introduced $\widehat{P}_{\ell,|m|}(\cos\te)$ defined by
\begin{align}
\label{def:Phat}
\widehat{P}_{\ell,|m|}(\cos\te) := \sin^{|m|}\te\frac{C^{(|m|+1/2)}_{\ell-|m|}(\cos\te)}{C^{(|m|+1/2)}_{\ell-|m|}(1)} \,,
\end{align} 
which is proportional to the associated Legendre function~\cite{AH}.

\medskip
We therefore conclude that
\begin{align}
\int_0^{2\pi} \frac{d\psi}{2\pi} \, e^{-im\psi} (\bsep.\bsk_2)^{s-n} (\bsep.\bsk_3)^n
=\frac{i^{s-{|m|}}}{2^{|m|}}
\binom{s-n}{|m|} \, k_2^{s-n}k_3^ne^{-im\chi} 
 \widehat{P}_{s-n,{|m|}}(\cos\te) \,.
\end{align}
It is then straightforward to derive Eqs.~\eqref{00s_helicity}-\eqref{A_123m} from this expression.

\section{Special functions}
\label{App:special_functions}

We summarize the definition and properties of special functions used in this paper.

\subsection{Jacobi polynomials}
\label{app:Jacobi}

The Jacobi polynomial is defined as
\begin{align}
\label{jacobi-def}
P^{(\al,\bt)}_n(t) = \frac{\Ga(\al+n+1)}{n!\,\Ga(\al+\bt+n+1)} \sum_{m=0}^n 
\binom{n}{m}
\frac{\Ga(\al+\bt+n+m+1)}{\Ga(\al+m+1)} 
\left( \frac{t-1}{2} \right)^m.
\end{align}
In general, the following Rodrigues' formula plays a central role:
\begin{align}
P^{(\al,\bt)}_n(t) = \frac{(-1)^n}{2^nn!} (1-t)^{-\al} (1+t)^{-\bt} 
\frac{d^n}{dt^n} \left\{ (1-t)^{\al+n} (1+t)^{\bt+n} \right\}.
\label{rodrigues}
\end{align}
The Jacobi polynomials are orthogonal in the sense that
\begin{align}
&{}
\int_{-1}^1 \! dt \, (1-t)^\al (1+t)^\bt  P^{(\al,\bt)}_m(t) P^{(\al,\bt)}_n(t) \nn\\
& \quad
= \frac{2^{\al+\bt+1}}{\al+\bt+2n+1} \frac{\Ga(\al+n+1) \Ga(\bt+n+1)}{n!\, \Ga(\al+\bt+n+1)}
\de_{mn}\,.
\label{jacobi-ortho}
\end{align}
The orthogonality for $m \neq n$ follows from the Rodrigues' formula \eqref{rodrigues}. 
The normalization factor can be found as a special case $\al=\ga$ of the integral formula (7.391.10, p.807, \cite{GR})
\begin{align}
&
\int_{-1}^1 \! dt \, (1-t)^\al (1+t)^\bt  P^{(\al,\bt)}_m(t) P^{(\ga,\bt)}_n(t) \nn\\
& 
= \frac{2^{\al+\bt+1}}{m!(n-m)!} 
\frac{\Ga(\ga+\bt+n+1+m) \Ga(\ga-\al+n-m) \Ga(\al+m+1) \Ga(\bt+n+1)}
{\Ga(\ga+\bt+n+1) \Ga(\ga-\al) \Ga(\al+\bt+m+n+2)}\,.
\label{jacobi-int}
\end{align}
We obtain Eq.~\eqref{jacobi-ortho} for $m=n$ by first setting $m=n$ to cancel $\Ga(\ga-\al)$ and then setting $\al=\ga$, whereas $m\neq n$ can be obtained by setting $\al=\ga$ first. 

\medskip
In this paper we use the Chebyshev and Gegenbauer polynomials, which are defined as special cases of the Jacobi polynomial. The Chebyshev polynomial is defined as
\begin{align}
\label{Chebyshev}
T_n(t) = \frac{n! \sqrt{\pi}}{\Ga(n+\frac{1}{2})} P_n^{(-\frac{1}{2}, -\frac{1}{2})}(t)\,,
\end{align}
which has the property $T_n(\cos\te)=\cos n\te$ and $T_0(t)=1$.
The Gegenbauer polynomial is defined as
\begin{align}
\label{Gegenbauer}
C^{(\al)}_n(t) = \frac{(2\al)_n}{(\al+\frac{1}{2})_n} P_n^{(\al-\frac{1}{2}, \al-\frac{1}{2})}(t)\,,
\end{align}
which satisfies $C_n^{(\al)}(1) = {(2\al)_n}/{n!}$.

\subsection{Bessel functions}
\label{App:triple-K}

The modified Bessel functions, $I$ and $K$, of the first and second kinds, respectively, are defined by
\begin{align}
I_\al(z) 
&= \sum_{n=0}^\infty \frac{(z/2)^{\al+2n}}{n!\,\Ga(\al+n+1)}\,, \\
K_\al(z)
&= \frac{\pi}{2\sin\pi\al}[I_{-\al}(z)-I_\al(z)]\,.
\end{align}
We call these functions the Bessel $I$ and $K$ functions in short. Also, the Bessel differential equation, the defining differential equation of these functions, is given by
\begin{align}
z^2 \frac{d^2f(z)}{dz^2} + z\frac{df(z)}{dz} - (z^2+\al^2)f(z) = 0\,,
\end{align}
which is useful to derive the bulk-to-boundary propagator $\cB_\nu(k;z)$ from the equation~\eqref{b-to-bd_def}.
A useful relation of the modified Bessel functions is
\begin{align}
\label{Bessel_useful}
\frac{1}{z}\frac{d}{dz}[z^\al K_\al(z)] = -z^{\al-1} K_{\al-1}(z)\,.
\end{align}
In the main text, we also use the asymptotic behavior around $z \sim \infty$, which is given by
\begin{align}
K_\al(z)\sim\sqrt{\frac{\pi}{2}}z^{-1/2}e^{-z}\,,
\quad
I_\al(z)\sim\frac{1}{\sqrt{2\pi}}z^{-1/2}e^z\,.
\end{align}

\bibliography{crossing}{}

\providecommand{\href}[2]{#2}\begingroup\raggedright\begin{thebibliography}{10}

\bibitem{Maldacena:1997re}
J.~M. Maldacena,  {\em {The Large N limit of superconformal field theories and
  supergravity}}, Int. J. Theor. Phys. {\bf 38} (1999) 1113--1133
  [\href{http://www.arXiv.org/abs/hep-th/9711200}{{\tt hep-th/9711200}}],
[Adv. Theor. Math. Phys.2,231(1998)].

\bibitem{Gubser:1998bc}
S.~S. Gubser, I.~R. Klebanov and A.~M. Polyakov,  {\em {Gauge theory
  correlators from noncritical string theory}}, Phys. Lett. {\bf B428} (1998)
  105--114
[\href{http://www.arXiv.org/abs/hep-th/9802109}{{\tt hep-th/9802109}}].

\bibitem{Witten:1998qj}
E.~Witten,  {\em {Anti-de Sitter space and holography}}, Adv. Theor. Math.
  Phys. {\bf 2} (1998) 253--291
[\href{http://www.arXiv.org/abs/hep-th/9802150}{{\tt hep-th/9802150}}].

\bibitem{Qualls:2015qjb}
J.~D. Qualls,  {\em {Lectures on Conformal Field Theory}},
\href{http://www.arXiv.org/abs/1511.04074}{{\tt 1511.04074}}.

\bibitem{Rychkov:2016iqz}
S.~Rychkov, {\em {EPFL Lectures on Conformal Field Theory in $D \geq 3$
  Dimensions}}.
\newblock SpringerBriefs in Physics.
2016.
\newblock

\bibitem{Simmons-Duffin:2016gjk}
D.~Simmons-Duffin,  {\em {The Conformal Bootstrap}}, in {\em {Proceedings,
  Theoretical Advanced Study Institute in Elementary Particle Physics: New
  Frontiers in Fields and Strings (TASI 2015): Boulder, CO, USA, June 1-26,
  2015}}, pp.~1--74.
\newblock 2017.
\newblock
\href{http://www.arXiv.org/abs/1602.07982}{{\tt 1602.07982}}.
\newblock

\bibitem{Penedones:2016voo}
J.~Penedones,  {\em {TASI lectures on AdS/CFT}}, in {\em {Proceedings,
  Theoretical Advanced Study Institute in Elementary Particle Physics: New
  Frontiers in Fields and Strings (TASI 2015): Boulder, CO, USA, June 1-26,
  2015}}, pp.~75--136.
\newblock 2017.
\newblock
\href{http://www.arXiv.org/abs/1608.04948}{{\tt 1608.04948}}.
\newblock

\bibitem{Maldacena:2002vr}
J.~M. Maldacena,  {\em {Non-Gaussian features of primordial fluctuations in
  single field inflationary models}}, JHEP {\bf 05} (2003) 013
[\href{http://www.arXiv.org/abs/astro-ph/0210603}{{\tt astro-ph/0210603}}].

\bibitem{Antoniadis:2011ib}
I.~Antoniadis, P.~O. Mazur and E.~Mottola,  {\em {Conformal Invariance, Dark
  Energy, and CMB Non-Gaussianity}}, JCAP {\bf 1209} (2012) 024
[\href{http://www.arXiv.org/abs/1103.4164}{{\tt 1103.4164}}].

\bibitem{Maldacena:2011nz}
J.~M. Maldacena and G.~L. Pimentel,  {\em {On graviton non-Gaussianities during
  inflation}}, JHEP {\bf 09} (2011) 045
[\href{http://www.arXiv.org/abs/1104.2846}{{\tt 1104.2846}}].

\bibitem{Creminelli:2012ed}
P.~Creminelli, J.~Nore\~{n}a and M.~Simonovi\'{c},  {\em {Conformal consistency
  relations for single-field inflation}}, JCAP {\bf 1207} (2012) 052
[\href{http://www.arXiv.org/abs/1203.4595}{{\tt 1203.4595}}].

\bibitem{Schalm:2012pi}
K.~Schalm, G.~Shiu and T.~van~der Aalst,  {\em {Consistency condition for
  inflation from (broken) conformal symmetry}}, JCAP {\bf 1303} (2013) 005
[\href{http://www.arXiv.org/abs/1211.2157}{{\tt 1211.2157}}].

\bibitem{Mata:2012bx}
I.~Mata, S.~Raju and S.~Trivedi,  {\em {CMB from CFT}}, JHEP {\bf 07} (2013)
  015
[\href{http://www.arXiv.org/abs/1211.5482}{{\tt 1211.5482}}].

\bibitem{McFadden:2013ria}
P.~McFadden,  {\em {On the power spectrum of inflationary cosmologies dual to a
  deformed CFT}}, JHEP {\bf 10} (2013) 071
[\href{http://www.arXiv.org/abs/1308.0331}{{\tt 1308.0331}}].

\bibitem{Ghosh:2014kba}
A.~Ghosh, N.~Kundu, S.~Raju and S.~P. Trivedi,  {\em {Conformal Invariance and
  the Four Point Scalar Correlator in Slow-Roll Inflation}}, JHEP {\bf 07}
  (2014) 011
[\href{http://www.arXiv.org/abs/1401.1426}{{\tt 1401.1426}}].

\bibitem{Bzowski:2012ih}
A.~Bzowski, P.~McFadden and K.~Skenderis,  {\em {Holography for inflation using
  conformal perturbation theory}}, JHEP {\bf 04} (2013) 047
[\href{http://www.arXiv.org/abs/1211.4550}{{\tt 1211.4550}}].

\bibitem{Kundu:2014gxa}
N.~Kundu, A.~Shukla and S.~P. Trivedi,  {\em {Constraints from Conformal
  Symmetry on the Three Point Scalar Correlator in Inflation}}, JHEP {\bf 04}
  (2015) 061
[\href{http://www.arXiv.org/abs/1410.2606}{{\tt 1410.2606}}].

\bibitem{Arkani-Hamed:2015bza}
N.~Arkani-Hamed and J.~Maldacena,  {\em {Cosmological Collider Physics}},
\href{http://www.arXiv.org/abs/1503.08043}{{\tt 1503.08043}}.

\bibitem{Kundu:2015xta}
N.~Kundu, A.~Shukla and S.~P. Trivedi,  {\em {Ward Identities for Scale and
  Special Conformal Transformations in Inflation}}, JHEP {\bf 01} (2016) 046
[\href{http://www.arXiv.org/abs/1507.06017}{{\tt 1507.06017}}].

\bibitem{Shukla:2016bnu}
A.~Shukla, S.~P. Trivedi and V.~Vishal,  {\em {Symmetry constraints in
  inflation, $\alpha$-vacua, and the three point function}}, JHEP {\bf 12}
  (2016) 102
[\href{http://www.arXiv.org/abs/1607.08636}{{\tt 1607.08636}}].

\bibitem{Isono:2016yyj}
H.~Isono, T.~Noumi, G.~Shiu, S.~S.~C. Wong and S.~Zhou,  {\em {Holographic
  non-Gaussianities in general single-field inflation}}, JHEP {\bf 12} (2016)
  028
[\href{http://www.arXiv.org/abs/1610.01258}{{\tt 1610.01258}}].

\bibitem{Ferrara:1973yt}
S.~Ferrara, A.~F. Grillo and R.~Gatto,  {\em {Tensor representations of
  conformal algebra and conformally covariant operator product expansion}},
  Annals Phys. {\bf 76} (1973)
161--188.

\bibitem{Belavin:1984vu}
A.~A. Belavin, A.~M. Polyakov and A.~B. Zamolodchikov,  {\em {Infinite
  Conformal Symmetry in Two-Dimensional Quantum Field Theory}}, Nucl. Phys.
  {\bf B241} (1984)
333--380.

\bibitem{Dolan:2000ut}
F.~A. Dolan and H.~Osborn,  {\em {Conformal four point functions and the
  operator product expansion}}, Nucl. Phys. {\bf B599} (2001) 459--496
[\href{http://www.arXiv.org/abs/hep-th/0011040}{{\tt hep-th/0011040}}].

\bibitem{Dolan:2003hv}
F.~A. Dolan and H.~Osborn,  {\em {Conformal partial waves and the operator
  product expansion}}, Nucl. Phys. {\bf B678} (2004) 491--507
[\href{http://www.arXiv.org/abs/hep-th/0309180}{{\tt hep-th/0309180}}].

\bibitem{Dolan:2011dv}
F.~A. Dolan and H.~Osborn,  {\em {Conformal Partial Waves: Further Mathematical
  Results}},
\href{http://www.arXiv.org/abs/1108.6194}{{\tt 1108.6194}}.

\bibitem{Penedones:2010ue}
J.~Penedones,  {\em {Writing CFT correlation functions as AdS scattering
  amplitudes}}, JHEP {\bf 03} (2011) 025
[\href{http://www.arXiv.org/abs/1011.1485}{{\tt 1011.1485}}].

\bibitem{Fitzpatrick:2011ia}
A.~L. Fitzpatrick, J.~Kaplan, J.~Penedones, S.~Raju and B.~C. van Rees,  {\em
  {A Natural Language for AdS/CFT Correlators}}, JHEP {\bf 11} (2011) 095
[\href{http://www.arXiv.org/abs/1107.1499}{{\tt 1107.1499}}].

\bibitem{Costa:2011dw}
M.~S. Costa, J.~Penedones, D.~Poland and S.~Rychkov,  {\em {Spinning Conformal
  Blocks}}, JHEP {\bf 11} (2011) 154
[\href{http://www.arXiv.org/abs/1109.6321}{{\tt 1109.6321}}].

\bibitem{ElShowk:2012ht}
S.~El-Showk, M.~F. Paulos, D.~Poland, S.~Rychkov, D.~Simmons-Duffin and
  A.~Vichi,  {\em {Solving the 3D Ising Model with the Conformal Bootstrap}},
  Phys. Rev. {\bf D86} (2012) 025022
[\href{http://www.arXiv.org/abs/1203.6064}{{\tt 1203.6064}}].

\bibitem{Hogervorst:2013sma}
M.~Hogervorst and S.~Rychkov,  {\em {Radial Coordinates for Conformal Blocks}},
  Phys. Rev. {\bf D87} (2013) 106004
[\href{http://www.arXiv.org/abs/1303.1111}{{\tt 1303.1111}}].

\bibitem{Hogervorst:2013kva}
M.~Hogervorst, H.~Osborn and S.~Rychkov,  {\em {Diagonal Limit for Conformal
  Blocks in $d$ Dimensions}}, JHEP {\bf 08} (2013) 014
[\href{http://www.arXiv.org/abs/1305.1321}{{\tt 1305.1321}}].

\bibitem{Kos:2013tga}
F.~Kos, D.~Poland and D.~Simmons-Duffin,  {\em {Bootstrapping the $O(N)$ vector
  models}}, JHEP {\bf 06} (2014) 091
[\href{http://www.arXiv.org/abs/1307.6856}{{\tt 1307.6856}}].

\bibitem{Gliozzi:2013ysa}
F.~Gliozzi,  {\em {More constraining conformal bootstrap}}, Phys. Rev. Lett.
  {\bf 111} (2013) 161602
[\href{http://www.arXiv.org/abs/1307.3111}{{\tt 1307.3111}}].

\bibitem{Gliozzi:2014jsa}
F.~Gliozzi and A.~Rago,  {\em {Critical exponents of the 3d Ising and related
  models from Conformal Bootstrap}}, JHEP {\bf 10} (2014) 042
[\href{http://www.arXiv.org/abs/1403.6003}{{\tt 1403.6003}}].

\bibitem{Hijano:2015zsa}
E.~Hijano, P.~Kraus, E.~Perlmutter and R.~Snively,  {\em {Witten Diagrams
  Revisited: The AdS Geometry of Conformal Blocks}}, JHEP {\bf 01} (2016) 146
[\href{http://www.arXiv.org/abs/1508.00501}{{\tt 1508.00501}}].

\bibitem{Isachenkov:2016gim}
M.~Isachenkov and V.~Schomerus,  {\em {Superintegrability of $d$-dimensional
  Conformal Blocks}}, Phys. Rev. Lett. {\bf 117} (2016), no.~7, 071602
[\href{http://www.arXiv.org/abs/1602.01858}{{\tt 1602.01858}}].

\bibitem{Alday:2016njk}
L.~F. Alday,  {\em {Large Spin Perturbation Theory for Conformal Field
  Theories}}, Phys. Rev. Lett. {\bf 119} (2017), no.~11, 111601
[\href{http://www.arXiv.org/abs/1611.01500}{{\tt 1611.01500}}].

\bibitem{Caron-Huot:2017vep}
S.~Caron-Huot,  {\em {Analyticity in Spin in Conformal Theories}}, JHEP {\bf
  09} (2017) 078
[\href{http://www.arXiv.org/abs/1703.00278}{{\tt 1703.00278}}].

\bibitem{Alday:2017zzv}
L.~F. Alday, J.~Henriksson and M.~van Loon,  {\em {Taming the
  $\epsilon$-expansion with Large Spin Perturbation Theory}},
\href{http://www.arXiv.org/abs/1712.02314}{{\tt 1712.02314}}.

\bibitem{Polyakov:1974gs}
A.~M. Polyakov,  {\em {Nonhamiltonian approach to conformal quantum field
  theory}}, Zh. Eksp. Teor. Fiz. {\bf 66} (1974) 23--42
[Sov. Phys. JETP39, 9 (1974)].

\bibitem{Sen:2015doa}
K.~Sen and A.~Sinha,  {\em {On critical exponents without Feynman diagrams}},
  J. Phys. {\bf A49} (2016), no.~44, 445401
[\href{http://www.arXiv.org/abs/1510.07770}{{\tt 1510.07770}}].

\bibitem{Gopakumar:2016wkt}
R.~Gopakumar, A.~Kaviraj, K.~Sen and A.~Sinha,  {\em {Conformal Bootstrap in
  Mellin Space}}, Phys. Rev. Lett. {\bf 118} (2017), no.~8, 081601
[\href{http://www.arXiv.org/abs/1609.00572}{{\tt 1609.00572}}].

\bibitem{Gopakumar:2016cpb}
R.~Gopakumar, A.~Kaviraj, K.~Sen and A.~Sinha,  {\em {A Mellin space approach
  to the conformal bootstrap}}, JHEP {\bf 05} (2017) 027
[\href{http://www.arXiv.org/abs/1611.08407}{{\tt 1611.08407}}].

\bibitem{crossing2}
H.~Isono, T.~Noumi and G.~Shiu,  {\em {Momentum space approach to crossing
  symmetric CFT correlators II: General spacetime dimension}}, {\it in
  preparation}.

\bibitem{Ferrara:1974nf}
S.~Ferrara, A.~F. Grillo, R.~Gatto and G.~Parisi,  {\em {Analyticity properties
  and asymptotic expansions of conformal covariant green's functions}}, Nuovo
  Cim. {\bf A19} (1974)
667--695.

\bibitem{Bzowski:2013sza}
A.~Bzowski, P.~McFadden and K.~Skenderis,  {\em {Implications of conformal
  invariance in momentum space}}, JHEP {\bf 03} (2014) 111
[\href{http://www.arXiv.org/abs/1304.7760}{{\tt 1304.7760}}].

\bibitem{Bzowski:2015pba}
A.~Bzowski, P.~McFadden and K.~Skenderis,  {\em {Scalar 3-point functions in
  CFT: renormalisation, beta functions and anomalies}}, JHEP {\bf 03} (2016)
  066
[\href{http://www.arXiv.org/abs/1510.08442}{{\tt 1510.08442}}].

\bibitem{Bzowski:2015yxv}
A.~Bzowski, P.~McFadden and K.~Skenderis,  {\em {Evaluation of conformal
  integrals}}, JHEP {\bf 02} (2016) 068
[\href{http://www.arXiv.org/abs/1511.02357}{{\tt 1511.02357}}].

\bibitem{Coriano:2012hd}
C.~Coriano, L.~Delle~Rose and M.~Serino,  {\em {Three and Four Point Functions
  of Stress Energy Tensors in D=3 for the Analysis of Cosmological
  Non-Gaussianities}}, JHEP {\bf 12} (2012) 090
[\href{http://www.arXiv.org/abs/1210.0136}{{\tt 1210.0136}}].

\bibitem{Coriano:2013jba}
C.~Coriano, L.~Delle~Rose, E.~Mottola and M.~Serino,  {\em {Solving the
  Conformal Constraints for Scalar Operators in Momentum Space and the
  Evaluation of Feynman's Master Integrals}}, JHEP {\bf 07} (2013) 011
[\href{http://www.arXiv.org/abs/1304.6944}{{\tt 1304.6944}}].

\bibitem{Bzowski:2017poo}
A.~Bzowski, P.~McFadden and K.~Skenderis,  {\em {Renormalised 3-point functions
  of stress tensors and conserved currents in CFT}},
\href{http://www.arXiv.org/abs/1711.09105}{{\tt 1711.09105}}.

\bibitem{Coriano:2018Feb}
C.~Coriano and M.~M. Maglio,  {\em {Exact Correlators from Conformal Ward
  Identities in Momentum Space and the Perturbative $TJJ$ Vertex}},
\href{http://www.arXiv.org/abs/1802.07675}{{\tt 1802.07675}}.

\bibitem{Elvang:2013cua}
H.~Elvang and Y.-t. Huang,  {\em {Scattering Amplitudes}},
\href{http://www.arXiv.org/abs/1308.1697}{{\tt 1308.1697}}.

\bibitem{Ferrara:1972xe}
S.~Ferrara and G.~Parisi,  {\em {Conformal covariant correlation functions}},
  Nucl. Phys. {\bf B42} (1972)
281--290.

\bibitem{Ferrara:1972ay}
S.~Ferrara, A.~F. Grillo and G.~Parisi,  {\em {Nonequivalence between conformal
  covariant wilson expansion in euclidean and minkowski space}}, Lett. Nuovo
  Cim. {\bf 5S2} (1972) 147--151
[Lett. Nuovo Cim. 5, 147 (1972)].

\bibitem{Ferrara:1972uq}
S.~Ferrara, A.~F. Grillo, G.~Parisi and R.~Gatto,  {\em {The shadow operator
  formalism for conformal algebra. vacuum expectation values and operator
  products}}, Lett. Nuovo Cim. {\bf 4S2} (1972) 115--120
[Lett. Nuovo Cim. 4, 115 (1972)].

\bibitem{Ferrara:1973vz}
S.~Ferrara, A.~F. Grillo, G.~Parisi and R.~Gatto,  {\em {Covariant expansion of
  the conformal four-point function}}, Nucl. Phys. {\bf B49} (1972) 77--98
[Erratum: Nucl. Phys. B53, 643 (1973)].

\bibitem{SimmonsDuffin:2012uy}
D.~Simmons-Duffin,  {\em {Projectors, Shadows, and Conformal Blocks}}, JHEP
  {\bf 04} (2014) 146
[\href{http://www.arXiv.org/abs/1204.3894}{{\tt 1204.3894}}].

\bibitem{positivity}
H.~Isono, T.~Noumi and G.~Shiu,  {\em {Positivity in AdS/CFT revisited}}, {\it
  in progress}.

\bibitem{Bargmann:1977gy}
V.~Bargmann and I.~T. Todorov,  {\em {Spaces of Analytic Functions on a Complex
  Cone as Carries for the Symmetric Tensor Representations of SO(N)}}, J. Math.
  Phys. {\bf 18} (1977)
1141--1148.

\bibitem{Costa:2011mg}
M.~S. Costa, J.~Penedones, D.~Poland and S.~Rychkov,  {\em {Spinning Conformal
  Correlators}}, JHEP {\bf 11} (2011) 071
[\href{http://www.arXiv.org/abs/1107.3554}{{\tt 1107.3554}}].

\bibitem{Dobrev:1977qv}
V.~K. Dobrev, G.~Mack, V.~B. Petkova, S.~G. Petrova and I.~T. Todorov,  {\em
  {Harmonic Analysis on the n-Dimensional Lorentz Group and Its Application to
  Conformal Quantum Field Theory}}, Lect. Notes Phys. {\bf 63} (1977)
1--280.

\bibitem{Chen:2009zp}
X.~Chen and Y.~Wang,  {\em {Quasi-Single Field Inflation and
  Non-Gaussianities}}, JCAP {\bf 1004} (2010) 027
[\href{http://www.arXiv.org/abs/0911.3380}{{\tt 0911.3380}}].

\bibitem{Baumann:2011nk}
D.~Baumann and D.~Green,  {\em {Signatures of Supersymmetry from the Early
  Universe}}, Phys. Rev. {\bf D85} (2012) 103520
[\href{http://www.arXiv.org/abs/1109.0292}{{\tt 1109.0292}}].

\bibitem{Noumi:2012vr}
T.~Noumi, M.~Yamaguchi and D.~Yokoyama,  {\em {Effective field theory approach
  to quasi-single field inflation and effects of heavy fields}}, JHEP {\bf 06}
  (2013) 051
[\href{http://www.arXiv.org/abs/1211.1624}{{\tt 1211.1624}}].

\bibitem{Costa:2018mcg}
M.~S. Costa and T.~Hansen,  {\em {AdS Weight Shifting Operators}},
\href{http://www.arXiv.org/abs/1805.01492}{{\tt 1805.01492}}.

\bibitem{Sleight:2016hyl}
C.~Sleight,  {\em {Interactions in Higher-Spin Gravity: a Holographic
  Perspective}}, J. Phys. {\bf A50} (2017), no.~38, 383001
[\href{http://www.arXiv.org/abs/1610.01318}{{\tt 1610.01318}}].

\bibitem{Sleight:2017fpc}
C.~Sleight and M.~Taronna,  {\em {Spinning Witten Diagrams}}, JHEP {\bf 06}
  (2017) 100
[\href{http://www.arXiv.org/abs/1702.08619}{{\tt 1702.08619}}].

\bibitem{Isono:2017grm}
H.~Isono,  {\em {On conformal correlators and blocks with spinors in general
  dimensions}}, Phys. Rev. {\bf D96} (2017), no.~6, 065011
[\href{http://www.arXiv.org/abs/1706.02835}{{\tt 1706.02835}}].

\bibitem{AH}
{K.~Atkinson and W.~Han},  {\em {Spherical Harmonics and Approximations on the
  Unit Sphere: An Introduction}}, Sptinger (2010).

\bibitem{GR}
I.~S. Gradshteyn and I.~M. Ryzhik,  {\em {Table of Integrals, Series and
  Products, Seventh Edition}}, Elsevier, Academic Press (2007).

\end{thebibliography}\endgroup
\bibliographystyle{utphys}

\end{document}